\documentclass[12pt, draftclsnofoot, onecolumn]{IEEEtran}
\usepackage{color}
\usepackage{float}
\usepackage{bm}
\usepackage{amsmath}
\usepackage{amssymb}
\usepackage{graphicx}
\usepackage{epstopdf}
\usepackage{cite}
\usepackage{amsfonts}
\usepackage{changepage}
\usepackage{subfigure}
\usepackage{algorithm}
\usepackage{algorithmicx}
\usepackage{multirow}
\usepackage{algpseudocode}
\usepackage{comment}
\usepackage{multirow}
\usepackage{booktabs}
\def\blue#1{{\color{black}#1}}
\def\bluenew#1{{\color{black}#1}}

\begin{document}
	
\title{Double-Sparsity Learning Based  Channel-and-Signal Estimation in  Massive MIMO with Generalized Spatial Modulation}
\author{Xiaoyan~Kuai,
	Xiaojun~Yuan,~\IEEEmembership{Senior~Member,~IEEE,}
	Wenjing~Yan,~Hang~Liu, and Ying Jun (Angela) Zhang,~\IEEEmembership{Senior~Member,~IEEE}
\thanks{X. Kuai, X. Yuan and W.~Yan are with the Center for Intelligent Networking and Communications,
the National Laboratory of Science and Technology on Communications, the University of Electronic Science and Technology of China, Chengdu 611731, China (e-mail: $\{$xy$\_$kuai, xjyuan$\}$@uestc.edu.cn).
}
\thanks{H. Liu and Y. J. Zhang are with the Department of Information Engineering,
The Chinese University of Hong Kong, Shatin, New Territories, Hong Kong (e-mail: \{lh117, yjzhang\}@ie.cuhk.edu.hk).}
\thanks{The work was partially presented in the IEEE International Conference on Communications in China (ICCC) \cite{xykuai19ICCC}.}
}

\markboth{}%
{Shell \MakeLowercase{\textit{et al.}}: Bare Demo of IEEEtran.cls for IEEE Journals}

\maketitle
	
\begin{abstract}
In this paper, we study joint antenna activity detection, channel estimation,
and multiuser detection for massive multiple-input multiple-output (MIMO) {systems}
with general spatial modulation (GSM). We first establish a double-sparsity massive MIMO
model by considering the channel sparsity of the massive MIMO channel and the
signal sparsity of GSM. Based on the double-sparsity model, we formulate a blind detection problem.
To solve the blind detection problem, we develop message-passing
based blind channel-and-signal estimation (BCSE)
algorithm.
The BCSE algorithm basically follows the affine sparse matrix factorization technique, but with critical
modifications to handle the double-sparsity property of the model.
We show that the BCSE algorithm significantly outperforms the existing blind and
training-based algorithms, and is able to closely approach the genie bounds (with either known channel or known signal).
In the BCSE algorithm, short pilots are employed to remove the phase and permutation ambiguities after sparse matrix factorization.
To utilize the short pilots more efficiently, we further develop the semi-blind channel-and-signal estimation (SBCSE) algorithm to incorporate the estimation of the
phase and permutation ambiguities into the iterative message-passing process.
We show that the SBCSE
algorithm substantially outperforms the counterpart algorithms including the BCSE algorithm in the short-pilot regime.
\end{abstract}

\IEEEpeerreviewmaketitle

\section{Introduction}
Wireless transceivers with large antenna arrays and powerful signal processing capabilities have been proposed to accommodate the exponential growth of data traffics. The new wireless infrastructures generally require a significant increase of energy consumption in establishing communication links \cite{li2011energy,bjornson2015optimal}. As such, the energy efficiency (EE) of wireless transmission has attracted intensive research interests in recent years.
Advanced technologies, such as massive multiple-input-multiple-output (MIMO) and spatial modulation (SM), have been developed to meet the EE requirement of next-generation wireless communication systems \cite{di2011spatial,naidoo2011spatial,rusek2013scaling}.

Massive MIMO with  spatial modulation  
is a new communication paradigm consisting of a multi-antenna base station (BS) and multiple multi-antenna users, where the
number of antennas at the BS is typically much greater than that at a user. In each time instance, every
user activates only one antenna  for signal transmission. Compared to conventional
modulation techniques, spatial modulation is a promising solution for multi-antenna
transmissions to reduce the power consumption, to relieve the burden of antenna synchronization,
and to mitigate the {inter-antenna} interference.  Recently, to achieve high spectrum efficiency,
generalized spatial modulation (GSM) has been proposed to allow the activation of multiple
antennas for simultaneous transmission of multiple independent symbols at each user \cite{narasimhan2015generalized}.

A key challenge for GSM-based massive MIMO is how to carry out antenna
activity detection, channel estimation, and mulituser detection at the BS.
Most existing work assumes perfect channel state information (CSI) (or assumes that the CSI can
be acquired from channel training in prior), and is focused on antenna activity detection and
multiuser detection. For example, the author in \cite{zheng2014low}  proposed
a two-step approach: In the first step, the indices of active antennas are estimated
 using the ordered nearest minimum square-error detector;
then in the second step, the signals are recovered based on the knowledge of the active antennas.
In contrast to the two-step approach, the authors in \cite{xu2013spatial} proposed
 a joint approach in which maximum likelihood (ML) detection is used to estimate both the indices
of active antennas and the signals transmitted by these active antennas. However, the ML-based method
suffers  prohibitively high computational complexity as the size of a MIMO system scales up.
Several low-complexity detectors with near-optimal performance was proposed in \cite{younis2013generalised,xu2013spatial}.

More recently, {researchers} have proposed to design GSM-based massive MIMO systems
by exploiting the signal sparsity inherent in spatial modulation \cite{liu2014denoising,narasimhan2014large,garcia2015low,wang2015multiuser}.
Specifically, the authors in \cite{liu2014denoising} and \cite{garcia2015low}
employed $l_1$ regularization based compressed sensing techniques \cite{donoho2006compressed}
for the recovery of sparse signals. In \cite{narasimhan2014large}, a message-passing algorithm was
developed for joint antenna activity detection and multiuser detection. In \cite{wang2015multiuser},
the authors proposed a generalized approximate message passing (GAMP) detector to deal with
quantized measurements and spatial correlation in a large-scale antenna array
at the BS.

The above mentioned approaches, however, have the following two limitations.
First, all these approaches assume that the CSI is either {\textit{a priori}}
known to the receiver or estimated in a separate training stage prior
to antenna activity detection and signal detection. In practice, the channel is unknown and the training-based
method  causes a significant pilot overhead when the MIMO size becomes large.
Second, the structure of the massive MIMO channel, such as the angular-domain sparsity and
the  correlation in antenna arrays, has not been fully exploited in the existing algorithms.

In this paper, we study the transceiver design of the GSM-based massive MIMO system to
address the above two limitations. We first establish a double-sparsity massive MIMO
model by considering the correlation between  transmit/receive antennas \cite{zhou2007experimental},
the clustered channel sparsity in the angular domain \cite{chen2017}, and
the signal sparsity inherent in GSM. \bluenew{Specifically, the received signal can be represented as $\mathbf{Y}=\mathbf{A}_{\textsf{R}}{\mathbf{G}}{\mathbf{A}}_{\textsf{T}}^H\mathbf{X}+\mathbf{N}$,
where $\mathbf{A}_{\textsf{R}}$ and ${\mathbf{A}}_{\textsf{T}}$ are steering vector matrices
characterizing the receive and transmit correlations, respectively,  ${\mathbf{G}}$ is a sparse angular-domain channel matrix,
$\mathbf{X}$ is a sparse signal matrix with GSM, and $\mathbf{N}$ is an ambient noise matrix.
With the knowledge of $\mathbf{A}_{\textsf{R}}$ and $\mathbf{A}_{\textsf{T}}$, the joint estimation of the sparse matrices $\mathbf{G}$ and $\mathbf{X}$ from $\mathbf{Y}$ is a bilinear recovery problem.
It seems that the parametric bilinear generalized approximate message passing (P-BiGAMP) algorithm \cite{parker2016parametric}
can be applied to this problem by vectorizing $\mathbf{Y}$, $\mathbf{G}$, and $\mathbf{X}$. However, we find that the P-BiGAMP algorithm does not work in our problem, probably because
the matrix product $\mathbf{A}_{\textsf{R}}\mathbf{G}\mathbf{A}_{\textsf{T}}^H\mathbf{X}$ here does not satisfy the requirement of random measurements by P-BiGAMP. To address this issue, we formulate a blind detection problem by absorbing ${\mathbf{A}}_{\textsf{T}}^H$ into the matrix either on the right or on the left in sparse matrix factorization.
To solve the blind detection problem,}
we develop a message passing based blind channel-and-signal estimation (BCSE) algorithm
that performs antenna activity activation, channel estimation, and user
detection simultaneously. \blue{We show that although the basic idea is borrowed from
the affine sparse matrix factorization (ASMF) algorithm developed in \cite{liu2018super},
new initialization and and re-initialization methods are necessary to ensure a good algorithm performance for the
considered double-sparsity model.}
We also show that our proposed scheme significantly outperforms the other
blind detection schemes \cite{zhang2017blind,ding2018sparsity} (that
exploit either the channel sparsity or the signal sparsity, but not both) and
the state-of-the-art training-based schemes for massive MIMO systems  with GSM \cite{wen2016bayes}\footnote{
The algorithm in \cite{wen2016bayes} is designed to handle low-resolution ADCs for massive MIMO systems.
With straightforward modifications, it can be applied to systems with high-resolution ADCs (as assumed in this paper).}.

Similar to the schemes developed in \cite{liu2018super,zhang2017blind,ding2018sparsity},
{sparse matrix factorization} suffers from the so-called phase and permutation ambiguities. {In the
BCSE algorithm,} reference symbols and {antenna} labels are used to eliminate
the phase and permutation ambiguities after matrix factorization.
Similar to {the} pilot signals in a training-based scheme, the reference symbols and
the antenna  labels are \textit{a priori} known by the receiver. Therefore,
they can be incorporated into the iterative detection process for performance enhancement,
rather than used for  compensation afterwards. As such, we develop a semi-blind channel-and-signal
 estimation (SBCSE) algorithm by treating the reference symbols and the antenna labels as
short pilots. Based on the framework of BCSE, we introduce two extra steps in
the SBCSE algorithm: We use the short pilots to eliminate the phase and permutation ambiguities in the
output of BCSE, and then use compressed sensing techniques to further refine the
channel estimate based on the structured sparsity of the massive MIMO channel.
Numerical results demonstrate that the proposed SBCSE algorithm substantially outperforms
the  state-of-the-art counterpart algorithms including the BCSE algorithm in the short-pilot regime.

To summarise, the main contributions of this paper are listed as follows:
\begin{itemize}
  \item To the best of our knowledge, this is the first work to consider joint antenna activity detection,
channel estimation, and multiuser detection based on the double-sparsity model for GSM-based massive MIMO systems.
We establish a comprehensive probability model to characterize the channel sparsity inherent in the massive MIMO
channel and the signal sparsity inherent in GSM, based on which the joint estimation problem is defined.
  \item We develop a message-passing based blind detection algorithm,
termed the BCSE algorithm, to efficiently exploit the channel sparsity and the signal sparsity.
We show that the BCSE algorithm significantly outperforms the existing blind and training-based algorithms, and
is able to closely approach the genie bounds (with either known channel or known signal).
  \item To  utilize the pilot signals (including the reference symbols and the user labels) more efficiently,
  we further develop a semi-blind detection algorithm, termed SBCSE.
  We show that the SBCSE algorithm substantially outperforms the counterpart algorithms
  including the BCSE algorithm in the short-pilot regime.
\end{itemize}

The rest of this paper is organized as follows. Section II describes the GSM-based massive MIMO systems. Section III and Section IV present the proposed blind and semi-blind channel-and-signal estimation algorithms, respectively.
We discuss the parameter learning and the complexity of the  proposed algorithms in Section V. Numerical results are presented in Section VI. Conclusions are drawn in Section VII.

{\it Notation:} Regular letters, lowercase bold letters, and capital bold letters represent scalars, vectors, and matrices, respectively.
 The superscripts $(\cdot)^{H}$, $(\cdot)^{*}$, $(\cdot)^{T}$, and $(\cdot)^{-1}$
represent the conjugate transpose, the conjugate, the transpose, and the inverse of a matrix, respectively;
$|\cdot|$ represents the cardinality of a set; $\|\cdot\|_0$ denotes the $l_0$ norm;
  $\|\cdot\|_{1}$ denotes the $l_1$ norm; $\|\cdot\|_{F}$ denotes the Frobenius norm.
$\rm diag\{\mathbf{a}\}$ represents the diagonal matrix with the diagonal {entries} specified by $\mathbf{a}$.
$\otimes$ denotes the Kronecker product and $\delta(\cdot)$ denotes the Dirac delta function.
$\mathbf{I}_N$ denotes the identity matrix of size $N \times N$. Some frequently used symbols are listed in the Table~\ref{tab00}.
\begin{table}
\small
\centering
\setlength{\abovecaptionskip}{0pt}
\label{tab00}
\caption{Frequently used symbols}
\begin{tabular}{|l|l|l|l|}
\hline
$n$ & transmit-antenna index  & $m$ & receive-antenna  index \\
$k$ & user  index &$t$ & time slot index \\
$K$& number of users & $T$ & coherence time \\
$M$& number of antennas at receiver & $M^{'}$&  number of AoA bins  \\
$N$&  number of antennas at each user & $N^{'}$&  number of  AoD bins   \\
$\bm \theta$&  collection of AoAs & $\bm \phi$&  collection of AoDs  \\
$\bm \vartheta$&  angular grid of AoAs & $\bm \varphi$& angular grid of AoDs \\
$\mathbf{a}_{\text R}$ &  steering vector at receiver &$\mathbf{a}_{\text T}$ &  steering vector at transmitter \\
$\mathbf{A}_{\textsf{R}}$ & steering-vector matrix at receiver & $\mathbf{A}_{\textsf{T}}$& steering-vector matrix at transmitter\\
$\lambda$ &  sparsity level of the channel $\mathbf G$ & $\rho$ &  sparsity level of the signal  \\
$q$ &  index of  AoA  bins  &$p$ &  index of AoD  bins    \\
\hline
\end{tabular}
\end{table}
\section{ System Model}
\label{section:sysmod}

\subsection{GSM-Based Massive MIMO Systems}
\begin{figure}
  \centering
  \includegraphics[width=4in]{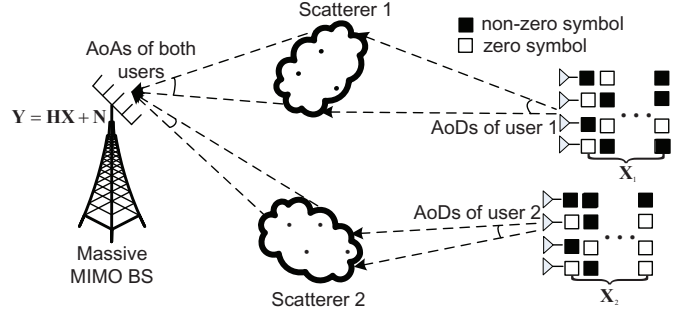}\\
  \caption{The illustration of  a massive MIMO system with  GSM for $K=2$ users.}\label{fig.GSM}
\end{figure}

We consider a multiple access system, in which $K$  users communicate with a single BS
 equipped with $M$ receive antennas.  $M$ is usually in the order of tens to hundreds.
Each user is equipped with $N$ transmit antennas and employs GSM \cite{narasimhan2015generalized,wang2015multiuser}.
 {That is}, at any time slot and for any user, each transmit antenna  either transmits a symbol
 taken from a modulation alphabet $\cal A$ or remains inactive (or in other words, transmits a zero symbol).\footnote{
We assume that the alphabet $\cal A$ is rotationally invariant for any rotation angle $\varpi \in \Omega$, i.e. ${\cal A}=e^{j\varpi}{\cal A}$, where $\Omega=\{\varpi_1,\cdots,\varpi_{|\Omega|}\}$
 is an angle set. For example, if the  quadrature phase shift keying (QPSK) modulation is involved, then $\Omega=\{0, \frac \pi 2, \pi, \frac {3\pi} {2}\}$.
The rotational invariance of $\cal A$ will be revisited when we discuss the ambiguity issue of  sparse matrix factorization.}
Specifically,
let $c_{k,n,t}$ be the indicator of the  activity state of antenna $n$ of user $k$ at time slot $t$, i.e.,
\begin{align}
c_{k,n,t}=\Bigg\{\begin{array}{cl}
           1, &    \text{antenna $n$ of}\ \text{user}  \ k \  \text{is active} \\
           0, & \text{otherwise}
         \end{array}
\end{align}
and $x_{k,n,t}$ be the symbol transmitted by antenna $n$ of  user $k$ at time slot $t$.
Note that $x_{k,n,t}\in \cal A$ if $c_{k,n,t}=1$, and $x_{k,n,t}=0$ if $c_{k,n,t}=0$.
We assume that $\{c_{k,n,t}\}$ are {independently} and identically distributed, and so are $\{x_{k,n,t}\}$.
In particular, each $x_{k,n,t}$ is independently drawn from the distribution of
\begin{align}\label{x.prior}
p(x)=\frac{\rho}{|\cal A|}\sum_{a\in \cal A}\delta(x-a)+(1-\rho)\delta(x),
\end{align}
where $\rho\in (0,1)$ is the signal sparsity level and $|\cal A|$ is the size of $\cal A$.
{Note that both $\rho$ and $\cal A$ are known to the receiver.}
We assume {that} the average power of $\cal A$ is normalized, i.e., $\frac{\sum_{a\in {\cal A}}|a|^2}{|\cal A|}=1$.  
Clearly, each antenna  transmits $H_{\rho}$ bits per time slot, where
\begin{align}
H_{\rho}=-(1-\rho)\log_2(1-\rho)-\rho\log_2\left(\frac{\rho}{|\cal A|}\right).
\end{align}
Denote by $\mathbf{x}_{k,t}=[x_{k,1,t},\cdots,x_{k,N,t}]^T$ the $k$-th
user's symbol vector at time slot $t$.
We stack all the symbol vectors from the $K$ users at time slot $t$ as
\begin{align}\label{sigal.users}
{\mathbf{x}}_t=[{\mathbf x}_{1,t}^T, {\mathbf x}_{2,t}^T,\cdots, \mathbf {x}_{K,t}^T]^T\in \mathbb{C}^{KN\times 1}.
\end{align}
Correspondingly, denote $\mathbf{H}\triangleq[\mathbf{h}_1,\cdots,\mathbf{h}_{KN}]\in \mathbb{C}^{M\times KN}$, where $\mathbf{h}_{(k-1)N+n}=[h_{1,(k-1)N+n},$ $h_{2,(k-1)N+n},\cdots,h_{M,(k-1)N+n}]^T\in \mathbb{C}^{M\times 1}$ is the
flat fading channel  coefficient vector  from  antenna $n$ of user $k$ to the BS.
 At time slot $t$, the received signal at the  BS  is given by
\begin{align}
{\mathbf y}_{t}=\mathbf{H}{\mathbf{x}}_t+\mathbf{n}_{t},
\end{align}
where  $\mathbf{n}_t$ is the AWGN noise following the complex  circularly symmetric Gaussian distribution with  mean zero and covariance $\sigma^2\mathbf{I}$
with $\sigma^2$ being the noise power and $\mathbf{I}$ being the identity matrix of an appropriate size.
We assume block fading with coherence time $T$, i.e., the channel remains unchanged for time duration of $T$.
Collecting all the received signals of $T$ successive  time slots, we  express the received signal at the BS  as
\begin{align}\label{input.output}
\mathbf{Y}=\mathbf{H}\mathbf{X}+\mathbf{N},
\end{align}
where $\mathbf{Y}=[\mathbf{y}_1,\mathbf{y}_2,\cdots,\mathbf{y}_T]$, $\mathbf{X}=[{\mathbf{x}}_1,\cdots,{\mathbf{x}}_T]$, and  $\mathbf{N}=[\mathbf{n}_1,\mathbf{n}_2,\cdots,\mathbf{n}_T]$.
The system model in \eqref{input.output} is illustrated in Fig.~\ref{fig.GSM}.
\subsection{{Angular-Domain} Channel Model}
We start with describing the channel representation in the {angular domain}.
During  coherence time $T$, the uplink channel from  user $k$ to the BS can be modelled as
\blue{
\begin{align}\label{channel.phy}
\mathbf{H}_k=\sum_{i=1}^{L_{k,c}}\sum_{j=1}^{L_{k,p}}\alpha_k(i,j)\mathbf{a}_{ {\textsf{R}} }(\theta_k(i,j))\mathbf{a}_{ \textsf{T}}^{H}(\phi_k(i,j)),
\end{align}
where {$L_{k,c}$} and {$L_{k,p}$} denote the number of scattering clusters and the number of physical paths in each cluster between user $k$
 and the BS, respectively; $\alpha_k(i,j)$ is the channel complex gain of path $j$ in
cluster $i$ for user $k$; $\mathbf{a}_{{\textsf{R}}}(\theta_k(i,j))$ and $\mathbf{a}_{{\textsf{T}}}(\phi_k(i,j))$
are the  steering vectors with $\theta_k$ being the angle of arrival (AoA) of BS and
 $\phi_k$ being the  angle of departure (AoD) of  user $k$, respectively.
For notational convenience, denote by $\bm{\theta}_k=\{\theta_k(i,j)\}_{\forall i,j}$ the collection of true AoAs and
{$\bm{\phi}_k=\{\phi_k(i,j)\}_{\forall i,j}$} the collection of true AoDs of user $k$.
In general, $\mathbf{a}_{ {\textsf{R}} }(\theta_k)$ and $\mathbf{a}_{ \textsf{T}}(\phi_k)$
are determined by the geometry of the antenna arrays at the BS and the users, respectively.}
For convenience of discussion, we focus on the case  that both the BS and  the users are equipped with  uniform linear arrays (ULAs).
Let $d_{\textsf{R}}$ and $d_{\textsf{T}}$ denote the antenna spacing at the BS and at each user, respectively.
Then,  the corresponding steering vectors are given by
\begin{align}\label{steering.vector}
\mathbf{a}_\textsf{R}(\theta_k)&=\frac {1}{\sqrt {M}}\left[1,e^{-j2\pi \frac{d_{\textsf{R}}\sin \theta_k}{\varrho}},\cdots,e^{-j2\pi \frac{(M-1)d_{\textsf{R}}\sin \theta_k}{\varrho}}\right]^{T}\notag \\
\mathbf{a}_\textsf{T}(\phi_{k})&=\frac {1}{\sqrt {N}}\left[1,e^{-j2\pi \frac{d_{\textsf{T}}\sin \phi_k}{\varrho}},\cdots,e^{-j2\pi \frac{(N-1)d_{\textsf{T}}\sin \phi_k}{\varrho}}\right]^{T},
\end{align}
where $\varrho$ is the wavelength of propagation,  $\theta_k\in \left(-\frac \pi 2,\frac \pi 2\right)$, and $\phi_k\in \left(-\frac \pi 2,\frac \pi 2\right)$.
 Note that the work in this paper can be readily extended to antenna arrays with other geometries, such as lens antenna arrays (LAA) \cite{liu2018super} and
2-dimensional antenna arrays \cite{dai2017fdd}.

The model parameters $\{{\bm \theta}_k,{\bm \phi}_k,{L_{k,c}},{L_{k,p}}\}$ are difficult to acquire in practice.
To avoid this difficulty, we
 introduce the so-called virtual
channel representation of \eqref{channel.phy}.
Let ${\bm {\vartheta}}=\{{\vartheta}_{q}\}{}_{q=1}^{M^{'}}$ be
a given grid that consists of $M^{'}$ discrete angular bins ranging from $-\frac \pi 2$ to $\frac \pi 2$.
Similarly, let ${\bm {\varphi}_k}=\{\varphi_{k,p}\}_{p=1}^{N^{'}}$ be a given grid that consists of $N^{'}$ discrete angular bins ranging from $-\frac \pi 2$ to $\frac \pi 2$. 
For sufficiently large $M^{'}$ and $N^{'}$, $\mathbf{H}_k$ can be well approximated
in the virtual AoA and AoD domain by
\begin{align}\label{H.vitural}
\mathbf{H}_k&=\sum_{q=1}^{M^{'}}\sum_{p=1}^{N^{'}}g_k(q,p){\mathbf a}_{\textsf{R}}({\vartheta}_{q}){\mathbf a}_{\textsf{T}}^{H}({\varphi}_{k,p})\notag \\
&=\mathbf{A}_\textsf{R}({\bm {\vartheta}})\mathbf{G}_k\mathbf{A}_{\textsf{T,k}}^{H}({\bm {\varphi}_k}),
\end{align}
where $\mathbf{G}_k \in \mathbb{C}^{M^{'} \times N^{'}}$ is the virtual angular-domain channel matrix with the $(q,p)$-th element given by $g_k(q,p)$,  $\mathbf{A}_{\textsf{R}}({\bm {\vartheta}})\triangleq[\mathbf{a}_{\textsf{R}}({\vartheta}_{1}),\cdots,\mathbf{a}_{\textsf{R}}({\vartheta}_{M^{'}})]\in \mathbb{C}^{M\times M^{'}}$,
and $\mathbf{A}_{\textsf{T,k}}({\bm {\varphi}}_{k})\triangleq[\mathbf{a}_{\textsf{T}}({\varphi}_{k,1}),\cdots,\mathbf{a}_{\textsf{T}}({\varphi}_{k,N^{'}})]$
 $\in \mathbb{C}^{N\times N^{'}}$.
With \eqref{H.vitural}, the received signal at the BS in \eqref{input.output} can be represented as
\begin{subequations}\label{input.output.sparse}
\begin{align}
\mathbf{Y}&=\mathbf{H}\mathbf{X}+\mathbf{N} \label{Y.H}  \\ 
&=\mathbf{A}_{\textsf{R}}({\bm {\vartheta}}){\mathbf{G}}{\mathbf{A}}_{\textsf{T}}^H({\bm {\varphi}})\mathbf{X}+\mathbf{N}, \label{Y.H.b}
\end{align}
\end{subequations}
where ${\mathbf{G}}=[{\mathbf{G}}_1,\cdots,{\mathbf{G}}_K]\in \mathbb{C}^{M^{'}\times KN^{'} }$,
and ${\mathbf{A}}_{\textsf{T}}({\bm {\varphi}})={\rm diag}\{\mathbf{A}_{\textsf{T,1}}({\bm {\varphi}}_1),\cdots,\mathbf{A}_{\textsf{T,K}}({\bm {\varphi}}_{K})\}\in \mathbb{C}^{KN\times KN^{'}}$.
Note that the electromagnetic signal of a user usually impinges upon or departs from an
antenna array in a limited number of angular bins, implying that  a large portion of the elements of $\mathbf{G}$ are zero, i.e., $\mathbf{G}$ is a sparse matrix.
Define the sparsity level of $\mathbf{G}$ as
\begin{align}
\lambda&=\frac{\|\mathbf{G}\|_0}{M^{'}N^{'}K},
\end{align}
where $\|\cdot\|_0$ denotes the $l_0$ norm. The system model in \eqref{Y.H.b} involves two sparse matrices $\mathbf{X}$
and $\mathbf{G}$, hence the name double-sparsity model.
\subsection{ Probability Model of $\mathbf{G}$}
\label{sec.prob.G}
Let $\mathbf{D}_k\in \mathbb{C}^{M^{'}\times N^{'}}$ be the channel support matrix of user $k$
and $d_{k,q,p}$ be the $(q,p)$-th {entry} of $\mathbf{D}_k$,
where $d_{k,q,p}=0$ (or 1) indicates that the corresponding {entry} $g_{k,q,p}$ of $\mathbf{G}_k$ is zero (or non-zero).
Following \cite{liu2018downlink}, we assume that the {entries} of $\mathbf{G}_k$ conditioned on $\mathbf{D}_k$ are independent
of each other, with the distribution given by
\blue{
 \begin{align}\label{channel.prior}
p(\mathbf{G}_k|\mathbf{D}_k)&=\prod_{q=1}^{M^{'}}\prod_{p=1}^{N^{'}}\left((1-d_{k,q,p})\delta(g_{k,q,p})+d_{k,q,p}{\cal C}{\cal N}\left(g_{k,q,p};0,v_{\rm pri}\right)\right),
  \end{align}
where $v_{\rm pri}$ is the variance of the non-zero {entries} of $\mathbf{G}_k$.} \blue{Note that
$v_{\rm pri}$ is determined by the large-scale fading of user $k$, and
is generally unknown to the receiver.}

Due to the limited number of scatterers in the propagation environment, the massive MIMO channels exhibit the property of clustered sparsity,
i.e., the non-zero {entries} of $\mathbf{G}$ usually gather in clusters, with each cluster corresponding to a scatterer, as illustrated in Fig.~\ref{fig.GSM}.
To exploit the clustered sparsity,
we shall introduce  Markov model to capture the
scattering structure at the transmitter and  {the receiver} \cite{liu2018downlink}.

\section{Blind Channel-and-Signal Estimation}
\label{blind.dection}
\subsection{\bluenew{Problem Formulation}}

{ Blind channel-and-signal estimation aims} to estimate $\mathbf{H}$
and $\mathbf{X}$ from the observed data matrix $\mathbf{Y}$ in \eqref{input.output.sparse}, without
using any pilot signals. This problem can be formulated as
\begin{align}\label{problem.HX}
(\hat{\mathbf{H}},\hat{\mathbf{X}})&=\arg\max_{\mathbf{H},\mathbf{X}}p(\mathbf{X},\mathbf{H}|\mathbf{Y}).
\end{align}
To solve \eqref{problem.HX}, our previous {work} proposed to factorize the noisy product $\mathbf{Y}$
by exploiting either the channel sparsity \cite{zhang2017blind,liu2018super} or the signal sparsity \cite{ding2018sparsity}.
Sparse matrix factorization techniques, such as  the
K-SVD algorithm \cite{aharon2006k}, the SPAMS algorithm \cite{mairal2010online}, the ER-SpUD algorithm \cite{spielman2012exact}, and the bilinear generalized approximate message passing (BiG-AMP) algorithm \cite{parker2014bilinear}, can be used to produce {the} estimates of $\mathbf{H}$ and $\mathbf{X}$ simultaneously.
It has been shown in \cite{zhang2017blind,liu2018super}, and  \cite{ding2018sparsity}
that the blind estimation approach suffers from phase and permutation ambiguities.
More {specifically},  denote by $\bm{\Sigma}$ a unitary diagonal matrix with the phases of the diagonal {entries} randomly selected from $ \Omega$ (see footnote
2 for the definition of $ \Omega$).
{Denote} by $\bm{\Pi}$ an arbitrary permutation matrix. The phase and permutation ambiguities are {due to} the fact
that if $(\hat {\mathbf{H}}, \hat{\mathbf{X}})$ is a solution to \eqref{problem.HX}, then $(\tilde{\mathbf{H}}=\hat {\mathbf{H}}\bm{\Pi}^{-1}\bm{\Sigma}^{-1}, \bm{\Sigma}\bm{\Pi}\hat{\mathbf{X}})$ is also a valid solution to \eqref{problem.HX}.
{The ambiguity issue has the following two consequences for  blind detection.}
On the one hand, the solution of \eqref{problem.HX} is not unique, and {thus}
extra resources (such as reference signals and user labels) are required to eliminate the ambiguities after performing
matrix factorization. On the other hand,
the existence of the ambiguities facilitates the design of efficient iterative algorithms to find  equally good solutions.
In fact, it has been shown in \cite{sun2017complete} that gradient-based iterative
algorithms can  find a globally optimal solution of the non-convex sparse matrix factorization problem,
provided that certain regularity conditions are satisfied.

The approaches in \cite{zhang2017blind,liu2018super}, and \cite{ding2018sparsity}, however, fail to
exploit the double-sparsity property of the model in \eqref{input.output.sparse}.
With this regard, we aim to design an efficient blind channel-and-signal estimation algorithm that
can simultaneously exploit the sparsity of both  channel matrix $\mathbf{G}$ and signal matrix $\mathbf{X}$.
{Rewrite} $\mathbf{Y}$ in \eqref{input.output.sparse} in its vectorized form as
{\begin{align}\label{vec.Y}
 {\rm vec}(\mathbf{Y})&=\sum_{q,p}\sum_{l}g_{q,p}\mathbf{z}_{q,p,l}x_l+{\rm vec}(\mathbf{N}),
\end{align}
where $g_{q,p}$ is the $(q,p)$-th element of $\mathbf{G}$, $x_{l}$  is the $l$-th element of ${\rm vec}(\mathbf{X})$, $\mathbf{z}_{q,p,l}\in \mathbb{C}^{MT}$ is the $l$-th column of $(\mathbf{I}_{T} \otimes \mathbf{B}_{q,p})$, where $\mathbf{B}_{q,p}\triangleq \mathbf{a}_{\textsf{R},q}({{\vartheta}_q}){\mathbf{a}}_{\textsf{T},p}^H({{\varphi}_p})\in \mathbb{C}^{M\times KN}$, and $\mathbf{a}_{\textsf{R},q}({ {\vartheta}_q})$ and ${\mathbf{a}}_{\textsf{T},p}^H({ {\varphi}_p})$
are respectively the $q$-th column of $\mathbf{A}_{\textsf{R}}({\bm {\vartheta}})$ and the $p$-th row
 of ${\mathbf{A}}_{\textsf{T}}^H({\bm {\varphi}})$.
 With the measurements in \eqref{vec.Y}, it seems that the joint estimation of $\mathbf{G}$ and $\mathbf{X}$ can be solved by
the parametric bilinear generalized approximate message passing (P-BiGAMP) algorithm \cite{parker2016parametric}.
However, through extensive simulations, we observe that the P-BiGAMP algorithm do not work in factorizing the sparse matrices $\mathbf{G}$ and $\mathbf{X}$.
For the P-BiGAMP algorithm, we conjecture  the main reason  as follows. Due to the existence of the known matrix $\mathbf{A}_{\textsf{T}}(\mathbf{\bm \varphi})$ between $\mathbf{G}$
 and $\mathbf{X}$ in \eqref{Y.H.b}, the aforementioned phase and permutation ambiguities \cite{zhang2017blind} no longer exist.
 {In other words}, the  solution to the factorization of $\mathbf{G}$ and $\mathbf{X}$ based on $\mathbf{Y}$ in \eqref{input.output.sparse}
is unique up to a scalar phase shift.\footnote{As a matter of fact, the factorization of $\mathbf{G}\mathbf{A}_{\textsf{T}}^{H}\mathbf{X}$
in \eqref{Y.H.b} generally suffers from scalar phase ambiguity, i.e., for any solution of $(\hat{\mathbf{G}},\hat {\mathbf{X}})$,
$(\hat{\mathbf{G}}e^{-j\varpi}, e^{j\varpi} \hat {\mathbf{X}})$ for $\varpi\in {\Omega}$ is still a valid solution,
where $\Omega$ consists of the rotation-invariant angles of $\cal A$.}
With such uniqueness of the solution, the P-BiGAMP algorithm is prone to be struck at a local optimum.}
{One way to avoid the above difficulty is to absorb} $\mathbf{A}_{\textsf{T}}(\bm{\varphi})$  into the matrix
either on the right or on the left in sparse matrix factorization. {Along this line}, we consider the following three
approaches for blind channel-and-signal estimation.
\begin{enumerate}
\item {We first consider a simplified DFT-based signal model.}
\blue{We  project the received signal
matrix to the angular domain by the inverse DFT unitary transform, i.e.
\begin{align}\label{Y.propose.1}
\mathbf{Y}^{'}&=\mathbf{F}^{H}\mathbf{Y} \notag \\
&=\mathbf{F}^{H}\mathbf{A}_{\textsf{R}}({\bm {\vartheta}}){\mathbf{G}}{\mathbf{A}}_{\textsf{T}}^H({\bm {\varphi}})\mathbf{X}+\mathbf{F}^{H}\mathbf{N} \notag \\
&=\mathbf{F}^{H}\mathbf{A}_{\textsf{R}}({\bm {\vartheta}})\mathbf{S}\mathbf{X}+\mathbf{N}^{'},
\end{align}
where $\mathbf{S}=\mathbf{G}\mathbf{A}_{\textsf{T}}^{H}(\bm{\varphi})$, and $\mathbf{N}^{'}=\mathbf{F}^{H}\mathbf{N}$.
Suppose that the channel AoAs are located on  a uniform sampling grid for virtual spatial angles, i.e.
\begin{align}\label{virtual.angle}
\sin \vartheta_q=\frac{q-1}{M^{'}}, \textrm{for}\; q=1,\cdots,M^{'}=M.
\end{align}
Substituting \eqref{virtual.angle} into \eqref{Y.propose.1}, we see that $\mathbf{A}_{\textsf{R}}(\bm {\vartheta})$ is the
normalized DFT matrix. Then $\mathbf{F}^{H}\mathbf{A}_{\textsf{R}}({\bm {\vartheta}})$ becomes  the identity matrix. 
Thus, we can estimate both $\mathbf{S}$ and $\mathbf{X}$ by directly factorizing  $\mathbf{Y}^{'}$,  which can be accomplished by using the BiGAMP algorithm. However, the AoAs are generally not on the grid in practice. This DFT-based method (in which the estimates of $\mathbf{S}$ and $\mathbf{X}$ are
obtained by treating $\mathbf{Y}^{'}$ modelled as $\mathbf{Y}^{'}=\mathbf{S}\mathbf{X}+\mathbf{N}^{'}$) always suffers  performance loss due to the unavoidable AoA mismatch, i.e. $\mathbf{F}^{H}\mathbf{A}_{\textsf{R}}({\bm {\vartheta}})$ is actually not the identity matrix.}

\item Alternatively, we  absorb $\mathbf{A}_{\textsf{T}}(\bm{\varphi})$ into the right  by
letting $\tilde{\mathbf X}\triangleq\mathbf{A}_{\textsf{T}}^{H}({\bm {\varphi}})\mathbf{X}\in \mathbb{C}^{KN^{'}\times T}$.
 Then,
\begin{align}\label{Y.propose.2}
\mathbf{Y}&=\mathbf{A}_{\textsf{R}}({\bm {\vartheta}}){\mathbf{G}}\tilde{\mathbf X}+\mathbf{N}.
\end{align}
 We  follow the affine sparse matrix factorization approach in \cite{liu2018super} to produce the estimates of $\mathbf{G}$ and $\tilde{\mathbf{X}}$
based on $\mathbf{Y}$ in \eqref{Y.propose.2} and {the sparsity of} $\mathbf{G}$
and $\tilde{\mathbf{X}}$. However, due to the mixing effect
of $\mathbf{A}_{\textsf{T}}(\bm{\varphi})$, the entries of $\tilde{\mathbf{X}}$ are generally not constrained on the alphabet
 $\cal A$. Such a loss of constellation constraints  leads to  performance degradation
in matrix factorization. 
  \item  To avoid the loss of constellation information, we absorb
$\mathbf{A}_{\textsf{T}}(\bm{\varphi})$ into the left matrix.  The system model is given by 
\begin{align}\label{Y.propose.3}
\mathbf{Y}&=\mathbf{A}_{\textsf{R}}({\bm {\vartheta}}){\mathbf S}\mathbf{X}+\mathbf{N}.
\end{align}
We  still follow the affine sparse matrix factorization approach in \cite{liu2018super} {to estimate}  $\mathbf{S}$ and $\mathbf{X}$ based on $\mathbf{Y}$. The only differences are that here
both $\mathbf{S}$ and $\mathbf{X}$  are sparse and that the entries of $\mathbf{X}$ are constrained on $\cal{A}$.
These properties  can be
exploited to  improve the performance of matrix factorization. 

\end{enumerate}

It is clear that all the models in \eqref{Y.propose.1}-\eqref{Y.propose.3}  involve the factorization
of two sparse matrices. In the following, we focus on the model in  \eqref{Y.propose.3} to
present the blind channel-and-signal estimation (BCSE) algorithm. The BCSE algorithm can be
applied to the models \eqref{Y.propose.1} and \eqref{Y.propose.2} with some minor modifications.
We will provide numerical evidences to show that the algorithm developed based on \eqref{Y.propose.3}
significantly outperform those based on \eqref{Y.propose.1} and \eqref{Y.propose.2}.

The  affine sparse matrix factorization (ASMF) problem {described above} can be formulated as
\begin{align}\label{post.prob}
( \hat {\mathbf S},\hat {\mathbf{X}})=\arg\max_{{\mathbf S},\mathbf{X}} p({\mathbf S},\mathbf{X}|\mathbf{Y};\bm{\psi}),
\end{align}
where $\bm{\psi}\triangleq \{{\bm \vartheta},\rho,\lambda_{{S}},p_{01}^{{S}},p_{10}^{{S}},v_{{S}},\sigma^2\}$, $\lambda_{{S}}$ is the sparsity level of $\mathbf{S}$,
$p_{01}^{{S}}$ and $p_{10}^{{S}}$ are the transition probabilities of the Markov chain characterizing the support structure of $\mathbf{S}$, and $v_{{S}}$ is the variance of the non-zero entries of $\mathbf{S}$.\footnote{Due to the mixing effect of $\mathbf{A}_{\textsf{T}}(\bm{\varphi})$,
the sparsity level $\lambda_{S}$ of $\mathbf{S}=\mathbf{G}\mathbf{A}_{\textsf{T}}^{H}(\bm{\varphi})$
generally satisfies $\lambda_{\textsf{T}}\geq \lambda_{{S}}\geq\lambda_{{G}}$. In addition, $p_{10}^S$
is not an independent parameter in $\bm{\psi}$, since $p_{10}^S$ can be obtained from
$\lambda_{{S}}$ and $p_{01}^S$ by using the equality $\lambda_{{S}}=\frac{p_{01}^{{S}}}{p_{01}^{{S}}+p_{10}^{{S}}}$.}
Here, the parameters in $\bm{\psi}$ are assumed to be known {when} solving \eqref{post.prob}.
The estimation of these parameters will be discussed later in Section V.
The problem in \eqref{post.prob} is generally difficult to solve.
In the following, we present a low-complexity approximate solution based on the
message passing principle.
\begin{figure}
  \centering
  \includegraphics[width=3.2in]{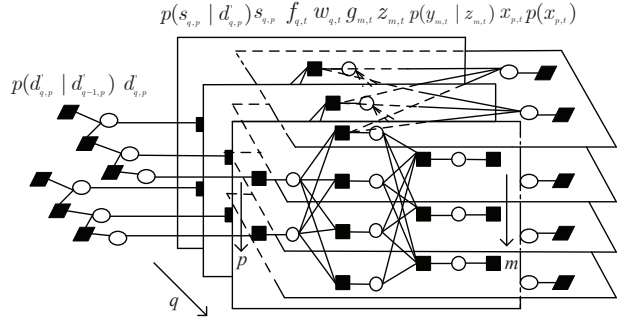}\\
  \caption{{An example of the factor graph for $M=3$, $T=4$, where hallow circles and black rectangles represent variable nodes and factor nodes, respectively, $f_{q,t}=\delta(w_{q,t}-\sum_{p=1}^{KN}s_{q,p}x_{p,t})$, $g_{m,t}=\delta(z_{m,t}-\sum_{q=1}^{M^{'}}A_{m,q}{w}_{q,t})$.} }\label{fig.factor.graph}
\end{figure}

\subsection{\bluenew{Factor Graph Representation}}

 To start with, we describe the factor graph representation of the probability distribution involved in \eqref{post.prob} as follows.
Recall that the entries of $\mathbf{X}_k$ are independently and uniformly drawn from the
distribution in \eqref{x.prior}.
Thus,
\begin{align}\label{prob.x}
p(\mathbf{X}_k)&=\prod_{n=1}^{N}\prod_{t=1}^{T}p(x_{k,n,t}).
\end{align}
Define $\mathbf{Z}\triangleq \mathbf{A}_{\textsf{R}}(\bm{\vartheta})\mathbf{S}\mathbf{X}\in \mathbb{C}^{M\times T}$.
Since $\mathbf{N}$ is an AWGN, we have
\begin{align}\label{prob.y}
p(\mathbf{Y}|\mathbf{\mathbf{Z}})&=\prod_{m=1}^{M}\prod_{t=1}^{T}{\cal C \cal N}(y_{m,t};z_{m,t},\sigma^2).
\end{align}
Due to the mixing effect of $\mathbf{A}_{\textsf{T}}(\bm \varphi)$,
the  Markovity of the support of $\mathbf{S}=\mathbf{G}\mathbf{A}_{\textsf{T}}^{H}(\bm\varphi)$
cannot be directly described by  the AoD and AoA random vectors introduced in Section II-C.
  Instead, let $\mathbf{D}^{'}$ denote the support of $\mathbf{S}$. We
use an independent Markov chain to describe the probability distribution of each column of $\mathbf{D}^{'}$, yielding
\begin{align}\label{markove.S}
p(\mathbf{D}^{'})&=\prod_{p=1}^{KN} \left(p(d_{1,p}^{'})\prod_{q=2}^{M^{'}} p(d_{q,p}^{'}|d_{q-1,p}^{'})\right),
\end{align}
where the transition probabilities are given by $p(d_{q,p}^{'}|d_{q-1,p}^{'}=0)=p_{10}^{{S}}\delta(d_{q,p}^{'}-1)+(1-p_{10}^{{S}})\delta(d_{q,p}^{'})$
and $p(d_{q,p}^{'}|d_{q-1,p}^{'}=1)=p_{01}^{{S}}\delta(d_{q,p}^{'})+(1-p_{01}^{{S}})\delta(d_{q,p}^{'}-1)$.
The initial $p(d_{1,p}^{'})$ is set as $p(d_{1,p}^{'})=\lambda_{S}\delta(d_{1,p}^{'}-1)+(1-\lambda_{S})\delta(d_{1,p}^{'})$.
Then, the joint
probability density distribution of  $( {\mathbf S},\mathbf{X})$  conditioning on $\mathbf {Y}$ is given by
\begin{align}\label{prob.joint}
&p({\mathbf S},\mathbf{X},\mathbf{Z},\mathbf{W}|\mathbf{Y})\propto p(\mathbf{Y}|\mathbf{Z})p(\mathbf{X})p({\mathbf S}|\mathbf{D}^{'})p(\mathbf{D}^{'})\delta(\mathbf{Z}-\mathbf{A}_{\textsf{R}}(\bm{\vartheta})\mathbf{W})\delta(\mathbf{W}-{\mathbf S}\mathbf X) \notag \\
&=\left(\prod_{m=1}^{M}\prod_{t=1}^{T}p(y_{m,t}|z_{m,t})\delta(z_{m,t}-\sum_{q=1}^{M^{'}}A_{m,q}{w}_{q,t})\delta(w_{q,t}-\sum_{p=1}^{KN}s_{q,p}{x}_{p,t})\right)\left(\prod_{p=1}^{KN}\prod_{t=1}^{T}p_{x_{p,t}}(x_{p,t})\right) \notag \\
&\quad   \times \left(\prod_{q=1}^{M^{'}}\prod_{p=1}^{KN}p(s_{q,p}|d_{q,p}^{'})\right)\!\!\left(\prod_{p=1}^{KN} \left(p(d_{1,p}^{'})\prod_{q=2}^{M^{'}} p(d_{q,p}^{'}|d_{q-1,p}^{'})\right)\!\right),
\end{align}
where $A_{m,q}$ is the $(m,q)$-th element of $\mathbf{A}_{\textsf{R}}(\bm{\vartheta})$. The factor graph representation of \eqref{prob.joint}
 is depicted in Fig.~\ref{fig.factor.graph}, where the variable nodes consist of  $\{z_{m,t}\}$, {$\{w_{q,t}\}$}, $\{{s}_{q,p}\}$, $\{{d}_{q,p}^{'}\}$, $\{x_{p,t}\}$,
and the check nodes consist of $\{f_{q,t}\}$, {$\{g_{m,t}\}$}, $\{p(x_{p,t})\}$,  $\{p({s}_{q,p})\}$, {$\{p(y_{m,t}|z_{m,t})\}$}, and  $\{p(d_{q,p}^{'}|d_{q-1,p}^{'})\}$.

\subsection{Blind Channel-and-Signal Estimation Algorithm}
\begin{algorithm}
\scriptsize
	\caption{\label{alg1}:  BCSE algorithm}
	\begin{algorithmic}[1]
		
		\State \textbf{Input:} received signal $\mathbf{Y}$, parameters  $\bm{\psi}\triangleq \{{\bm \vartheta},\rho,\lambda_{{S}},p_{01}^{{S}},p_{10}^{{S}},v_{{S}},\sigma^2\}$, prior distributions $p(x_{p,t})$ and $p(s_{q,p})$.
		
		
		\State\textbf{Initialization:} ${\hat s}_{q,p}(1)={\hat w}_{q,t}(1)=0$, $ v_{q,p}^{s}(1)=v_{q,t}^{w}(1)=v_{\rm max}$, $\hat {x}_{p,t}(1)$  randomly chosen from ${\cal A}$, $v_{p,t}^x(1)=v_{\rm max}$, $\hat {\tau}_{m,t}(0)=\hat{\alpha}_{q,t}(0)=0$, $\forall q,p,t$
		
		\For{$i=1,2,3,\cdots, I_{\rm max}$ } \quad\quad\quad\quad{$ \textbf{\%}$} \textbf{outer iteration}
\For{$j=1,2,3,\cdots, J_{\rm max}$ }\quad\quad\quad\quad{$ \textbf{\%}$} \textbf{inner iteration}
         \State $\forall m,t$: $v_{m,t}^{u}(j)=\sum_{q=1}^{M^{'}}|{A}_{m,q}|^2v_{q,t}^{w}(j)$;\;$\hat u_{m,t}(j)=\sum_{q=1}^{M^{'}}{A}_{m,q}\hat w_{q,t}(j)-v_{m,t}^{u}(j)\hat \tau_{m,t}(j-1)$
         \State $\forall m,t$: $v_{m,t}^z(j)=\frac{v_{m,t}^{u}(j)\sigma^2}{v_{m,t}^{u}(j)+\sigma^2}$;\;$\hat z_{m,t}(j)=\frac{\hat u_{m,t}(i)\sigma^2+y_{m,t}v_{m,t}^{u}(j)}{v_{m,t}^{u}(j)+\sigma^2}$
\State $\forall m,t$: $v_{m,t}^{\tau}(j)=\frac{v_{m,t}^{u}(j)-v_{m,t}^z(j)}{|v_{m,t}^u(j)|^2}$;\;$\hat {\tau}_{m,t}(j)=\frac{\hat z_{m,t}(j)-\hat u_{m,t}(j)}{v_{m,t}^{u}(j)}$
\State $\forall q,t$: $v_{q,t}^{\zeta} (j)=\left(\sum_{m=1}^{M}|A_{m,q}|^2v_{m,t}^{\tau}(j)\right)^{-1}$;\;${\hat \zeta}_{q,t}(j)=\hat w_{q,t}(j)+v_{q,t}^{\zeta} (j)\sum_{m=1}^{M}A_{m,q}{\hat \tau}_{m,t}(j)$
\State $\forall q,t$: $v_{q,t}^{\eta}(j)=\sum_{p=1}^{KN}|\hat s_{q,p}(j)|^2v_{p,t}^x(j)+v_{q,p}^s(j)|{\hat x}_{p,t}(j)|^2+v_{q,p}^s(j)v_{p,t}^x(j)$
\State $\forall q,t$: ${\bar \eta}_{q,t}(j)=\sum_{p=1}^{KN}\hat s_{q,p}(j)\hat x_{p,t}(j)$
\State $\forall q,t$: ${\hat \eta}_{q,t}(j)={\bar \eta}_{q,t}(j)-{\hat \alpha}_{q,t}(j-1)\left(v_{q,p}^s(j)|{\hat x}_{p,t}(j)|^2+|{\hat s}_{q,p}(j)|^2v_{p,t}^x(j)\right)$
\State $\forall q,t$: $\hat w_{q,t}(j)=v_{q,t}^w(j)\left(\frac{{\hat \zeta}_{q,t}(j)}{v_{q,t}^{\zeta}(j)}+\frac{{\hat \eta}_{q,t}(j)}{v_{q,t}^{\eta}(j)}\right)$;\;$v_{q,t}^w(j)=\frac{v_{q,t}^{\eta}(j)v_{q,t}^{\zeta}(j)}{v_{q,t}^{\eta}(j)+v_{q,t}^{\zeta}(j)}$
\State $\forall q,t$: $v_{q,t}^{\alpha}(j)=\frac{v_{q,t}^{w}(j)-v_{q,t}^{\eta}(j)}{|v_{q,t}^{\eta}(j)|^2}$;\;$\hat {\alpha}_{q,t}(j)=\frac{\hat w_{q,t}(j)-\hat \eta_{q,t}(j)}{v_{q,t}^{\eta}(j)}$
\State $\forall p,t$: $v_{p,t}^{r}(j)=\left(\sum_{q=1}^{M^{'}}|{\hat s}_{q,p}(j)|^2v_{q,t}^{\alpha}(j)\right)^{-1}$
\State $\forall p,t$: $\hat r_{p,t}(j)=\hat x_{p,t}(j)\left(1-v_{p,t}^r(j)\sum_{q=1}^{M^{'}}v_{q,p}^{s}(j)v_{q,t}^{\alpha}(j)\right)+v_{p,t}^{r}(j)\sum_{q=1}^{M^{'}}\hat s_{q,p}^{*}(j){\hat \alpha}_{q,t}(j)$
\State $\forall q,p$: $v_{q,p}^{\varsigma}(j)=\left(\sum_{t=1}^{T}|{\hat x}_{p,t}(j)|^2v_{q,t}^{\alpha}(j)\right)^{-1}$
\State $\forall q,p$: $\hat {\varsigma}_{q,p}(j)=\hat s_{q,p}(j)\left(1-v_{q,p}^{\varsigma}(j)\sum_{t=1}^{T}v_{p,t}^x(j)v_{q,t}^{\alpha }\right)+v_{q,p}^{\varsigma}(j)\sum_{t=1}^{T}{\hat x}_{p,t}^{*}(j){\hat \alpha}_{q,t}(j)$
\State $\forall q,p$: $\chi_{q,p}^{out}=\left(1+\frac{{\cal C\cal N}(0;\hat {\varsigma}_{q,p}(j),v_{q,p}^{\varsigma}(j))}{\int{\cal C\cal N}(s_{q,p};0,\sigma^2){\cal C\cal N}(s_{q,p};\hat {\varsigma}_{q,p}(j),v_{q,p}^{\varsigma}(j)) }\right)^{-1}$
\For {$q=2,\cdots,M^{'}$ }
\State$\lambda_{q,p}^{f}=\frac{p_{01}^{{S}}(1-\chi_{q-1,p}^{out})(1-\lambda_{q-1,p}^{f})+p_{11}^{{S}}\chi_{q-1,p}^{out}\lambda_{q-1,p}^{f}}{(1-\chi_{q-1,p}^{out})(1-\lambda_{q-1,p}^{f})+\chi_{q-1,p}^{out}\lambda_{q-1,p}^{f}}$, where $\lambda_{1,p}^f=\lambda_{{S}}$
\EndFor
\For {$q=M^{'}-1,\cdots,1$ }
\State$\lambda_{q,p}^{b}=\frac{p_{10}^{{S}}(1-\chi_{q+1,p}^{out})(1-\lambda_{q+1,p}^{b})+(1-p_{10}^{{S}})\chi_{q+1,p}^{out}\lambda_{q+1,p}^{b}}{(1-\chi_{q+1,p}^{out})(1-\lambda_{q+1,p}^{b})+\pi_{q+1,p}^{out}\lambda_{q+1,p}^{b}}$, where $\lambda_{M^{'},p}^b=\frac 1 2$
\EndFor
\State $\chi_{q,p}^{in}=\frac{\lambda_{q,p}^f\lambda_{q,p}^b}{(1-\lambda_{q,p}^f)(1-\lambda_{q,p}^b)+\lambda_{q,p}^f\lambda_{q,p}^b}$
\State $\forall p,t$: $\hat x_{p,t}(j+1)={\rm E}_{\tilde p(x_{p,t})}[x_{p,t}|\hat r_{p,t}(j),v_{p,t}^r(j),\bm{\psi}]$
\State $\forall p,t$: $v_{p,t}^{x}(j+1)={\rm E}_{\tilde p(x_{p,t})}[|x_{p,t}-\hat x_{p,t}(j+1)|^2|\hat r_{p,t}(j),v_{p,t}^r(j),\bm{\psi}]$
\State $\forall q,p$: $\hat s_{q,p}(j+1)={\rm E}_{\tilde p(s_{q,p})}[s_{q,p}|\hat {\varsigma}_{q,p}(j),v_{q,p}^{\varsigma}(j),\pi_{q,p}^{in},\bm{\psi}]$
\State $\forall q,p$: $v_{q,p}^{s}(j+1)={\rm E}_{\tilde p(s_{q,p})}[|s_{q,p}-\hat s_{q,p}(j+1)|^2|\hat {\varsigma}_{q,p}(j),v_{q,p}^{\varsigma}(j),\pi_{q,p}^{in},\bm{\psi}]$
        \EndFor
\State  \textbf{Re-initialize} $\hat x_{p,t}(1)$, $v_{p,t}^x(1)$, $\hat s_{q,p}(1)$,  and $v_{q,p}^s(1)$, $\forall p,q,t$
        \EndFor
\State $\forall p,t$: $\hat x_{p,t}=\hat x_{p,t}(j+1){ x}_{ref}/{\hat x}_{p,1}(j+1)$
\State $\forall q,p$: $\hat s_{q,p}=\hat s_{q,p}(j+1){\hat x}_{p,1}(j+1)/{ x}_{ref}$
\State\textbf{Output:} $\hat {\mathbf X}$, $\hat {\mathbf S}$	
	\end{algorithmic}
\end{algorithm}

The inference problem in Fig.~\ref{fig.factor.graph} can be solved by  the affine sparse matrix factorization method in \cite{liu2018super}.
The resulting BCSE algorithm is summarized in Algorithm 1. Most derivation details of Algorithm 1 can be found in \cite{liu2018super},
and thus are omitted for brevity. Here, we focus on the {difference} and provide a brief explanation of the algorithm based on
message passing over the factor graph in Fig.~\ref{fig.factor.graph}. \blue{The other differences  about new initialization and
re-initialization methods are presented in Section D.}

In  lines 5-8 of Algorithm 1, we adopt the approximate message passing principle \cite{rangan2011generalized} to
 calculate the messages from nodes $\{z_{m,t}\}$ to nodes $\{f_{q,t}\}$ based on the observations $\{y_{m,t}\}$.
 More specifically, in line 5, the messages from  nodes $\{f_{q,t}\}$ are cumulated
 to obtain an estimate of $\mathbf{A}_{\textsf{R}}(\bm \vartheta)\mathbf{W}$ with means $\{\hat {u}_{m,t}\}$ and variances
$\{v_{m,t}^u\}$, where $ \mathbf{W}\triangleq\mathbf{S}\mathbf{X}$ and ``Onsager" correction
is applied to generate the means $\{\hat {u}_{m,t}\}$.  Line 6 computes the means $\{\hat {z}_{m,t}\}$ and variances $\{v_{m,t}^z\}$
based on  $\{\hat {u}_{m,t}\}$, $\{v_{m,t}^u\}$, and  observations $\{y_{m,t}\}$.
In line 7, we compute the scaled residuals $\{\hat{\tau}_{m,t}\}$ and inverse-residual-variances $\{v_{m,t}^{\tau}\}$.
Then in line 8, the messages from nodes $\{z_{m,t}\}$ to nodes $\{f_{q,t}\}$
are combined to compute a estimate $ \{{w}_{q,t}\}$ with means $\{\hat{\zeta}_{q,t}\}$ and variances $\{v_{q,t}^{\zeta}\}$. In lines 9-11, the messages from nodes $\{s_{q,p}\}$, and $\{x_{p,t}\}$
are cumulated to obtain an estimate of $\mathbf{S}\mathbf{X}$ with means $\{\hat{\eta}_{q,t}\}$
and variance $\{v_{q,t}^{\eta}\}$. {Similar to line 7, line 13}  computes  scaled residuals and  inverse-residual-variances.
In lines 14-15, the messages from $\{f_{q,t}\}$ to node $\{x_{k,t}\}$ are combined to
compute  estimates of $\{x_{k,t}\}$, with means $\{\hat {r}_{k,t}\}$ and  variances $\{v_{k,t}^r\}$.
 Then in lines 26-27, each pair of $\hat r_{p,t}$
and variance $v_{p,t}^x$  are merged with the prior distribution $p(x_{p,t})$ to produce the posterior
mean $\hat x_{p,t}$ and variance $v_{p,t}^x$, where the expectation is taken with respect to
\begin{align}\label{post.x}
\tilde {p}(x_{p,t})=\frac{p(x_{p,t}){\cal C\cal N}(x_{p,t};{\hat r}_{p,t},v_{p,t}^r)}{\int p(x_{p,t}){\cal C\cal N}(x_{p,t};{\hat r}_{p,t},v_{p,t}^r)}.
\end{align}
We plug $p(x_{p,t})$ in \eqref{x.prior} into \eqref{post.x}, yielding
   \begin{align}\label{post.x.prob}
 & \tilde p({x}_{p,t})\!\! =\!\! \frac{\frac{\rho}{|\cal A|}\sum_{a\in \cal A}\delta(x_{p,t}-a){\cal C\cal N}\left(x_{p,t};\hat r_{p,t},v_{p,t}^r\right)\!\!+\! (1-\rho)\delta (x_{p,t}){\cal C\cal N}\left(x_{p,t};\hat r_{p,t},v_{p,t}^r\right)}{\int\frac{\rho}{|\cal A|}\sum_{a\in\cal A}\delta(x_{p,t}-a){\cal C\cal N}\left(x_{p,t};\hat r_{p,t},v_{p,t}^r\right)\!\!+\! (1-\rho)\delta (x_{p,t}){\cal C\cal N}\left(x_{p,t};\hat r_{p,t},v_{p,t}^r\right)}.
   \end{align}
Similar calculations are performed for $\{s_{q,p}\}$ in lines 16-17 and lines 28-29.
In line 18, we compute the messages from nodes $\{p(s_{q,p}|d_{q,p}^{'})\}$ to nodes $\{d_{q,p}^{'}\}$ given by $\{\chi_{q,p}^{out}d_{q,p}^{'}+(1-\chi_{q,p}^{out})(1-d_{q,p}^{'})\}$.
In lines 19-24,  forward and backward message passing \cite{chen2017} is applied.
In line 25, the messages from nodes $\{d_{q,p}^{'}\}$   to nodes $\{p(s_{q,p}|d_{q,p}^{'})\}$ are calculated by $\{\chi_{q,p}^{in}d_{q,p}^{'}+(1-\chi_{q,p}^{in})(1-d_{q,p}^{'})\}$.
In lines 28-29,  the expectation is taken with respect to
the distribution
\begin{align}\label{post.s}
\tilde p(s_{q,p})=\frac{p(s_{q,p}){\cal C\cal N}(s_{q,p};{\hat {\varsigma}}_{q,p},v_{q,p}^{\varsigma})}{\int p(s_{q,p}){\cal C\cal N}(s_{q,p};{\hat {\varsigma}}_{q,p},v_{q,p}^{\varsigma})},
\end{align}
with $p(s_{q,p})=(1-\chi_{q,p}^{in})\delta(s_{q,p})+\chi_{q,p}^{in}{\cal C\cal N}(s_{q,p};0,v_{{S}})$ is the message from node $p(s_{q,p}|d_{q,p}^{'})$
to node $s_{q,p}$. 
 Finally, since the first column of $\mathbf{X}$ is set as $x_{1,1}=x_{2,1}=\cdots=x_{KN,1}=x_{ref}$ (where $x_{ref}$ is a reference symbol),  the phase ambiguity can be eliminated in lines 33-34.\footnote{Then the permutation  ambiguity can
be eliminated by inserting an antenna label in each row of $\mathbf{X}$. To
assign a unique label for each transmit antenna, we need $\lceil \log_2KN\rceil$ bits, or equivalently, $\lceil \log_{|\cal A|}KN\rceil$ symbols,
for each label. }

\subsection{Initialization and Re-initialization}

The matrix factorization problem in \eqref{post.prob} is non-convex, and the iterative algorithm
described in Subsection C is prone to  {get stuck} at a local optimum.
 To alleviate this issue, we introduce  inner and outer iterations in the algorithm (following \cite{zhang2017blind,liu2018super},
and \cite{ding2018sparsity}), where multiple random initializations in the outer iteration and re-initializations in the inner iteration are employed
to avoid local optima.
For random initializations at the outer iteration, {the means of the initial signals}  are set as  symbols randomly chosen from $\cal A$,
and {the means of the initial channels} are set to zero.
The variances of the signal and the channel are initialized to $v_{\rm max}$.

We now discuss the re-initialization at the inner iteration. In \cite{zhang2017blind,liu2018super,ding2018sparsity},  either the channel or the signal is sparse,
but not both. The re-initialization  at the inner iteration is to reset the means and variances of the sparse variables (the channel or the signal),
while keeping the means and variances of non-sparse variables {the same as} the previous round of inner iteration.
This re-initialization method cannot be {directly applied} in Algorithm 1 since here both the channel matrix $\mathbf{S}$
 and the signal matrix $\mathbf{X}$ are sparse. As such, we explore the following five candidate methods for  re-initialization:
\begin{itemize}
  \item [(i)] {\it Reset channel mean and variance:} The first method only resets the  channel variables, i.e.,
line 31 of Algorithm 1 is replaced by ``$\hat x_{p,t}(1)=\hat x_{p,t}(j+1)$, $ v_{p,t}^x(1)=\hat v_{p,t}^x(j+1)$, $\hat s_{q,p}(1)=0$, and $ v_{q,p}^s(1)=v_{\rm max}$, $\forall p, t,q$.''
  \item [(ii)]{\it Reset channel mean, channel variance, and signal variance:} Line 31 of Algorithm 1 is replaced by ``$\hat x_{p,t}(1)=\hat x_{p,t}(j+1)$, $ v_{p,t}^x(1)=v_{\rm max}$, $\hat s_{q,p}(1)=0$, and $ v_{q,p}^s(1)=v_{\rm max}$, $\forall p, t,q$.''
  \item [(iii)]{\it Reset signal mean and variance:} Line 31 of Algorithm 1 is replaced by ``$\hat x_{p,t}(1)$ is randomly chosen
 from $\cal A$, $ v_{p,t}^x(1)=v_{\rm max}$, $\hat s_{q,p}(1)=\hat s_{q,p}(j+1)$, and $ v_{q,p}^s(1)=\hat v_{q,p}^s(j+1)$, $\forall p, t,q$.''
  \item [(iv)]{\it Reset signal mean, signal variance, and channel variance:} Line 31 of Algorithm 1 is replaced by ``$\hat x_{p,t}(1)$ is randomly chosen
 from $\cal A$, $ v_{p,t}^x(1)=v_{\rm max}$, $\hat s_{q,p}(1)=\hat s_{q,p}(j+1)$ and $ v_{q,p}^s(1)=v_{\rm max}$, $\forall p, t,q$.''
  \item [(v)]{\it Reset channel variance and signal variance:} Line 31 of Algorithm 1 is replaced by ``$\hat x_{p,t}(1)=\hat x_{p,t}(j+1)$, $\hat s_{q,p}(1)=\hat s_{q,p}(j+1)$,
and $ v_{p,t}^x(1)= v_{q,p}^s(1)=v_{\rm max}$, $\forall p, t,q$.''
\end{itemize}
In Section~VI, we present  simulation results to compare the above five methods. We
show by numerical simulations that the last method has the best performance among the five choices.

\section{Semi-Blind Channel-and-Signal Estimation}
In this section, we develop a semi-blind channel-and-signal estimation (SBCSE) algorithm,
as inspired by the following two reasons. First, as discussed in
Section~\ref{blind.dection}, blind channel-and-signal estimation suffers from the  phase and permutation
ambiguities inherent in matrix factorization.
One reference symbol and an antenna label
are inserted into each row of $\mathbf{X}_k$, $k=1,\cdots,K$ to eliminate the phase and permutation
ambiguities. Yet, as the reference symbols and antenna labels (similar to pilots) are \textit{a priori}
known by the receiver, such knowledge can be integrated into the iterative process of sparse
 matrix factorization  to improve the reliability of blind detection.
Second, recall that $\mathbf{S}=\mathbf{G}\mathbf{A}_{\textsf{T}}^H(\bm{\varphi})$,
where $\mathbf{A}_{\textsf{T}}^H(\bm{\varphi})$ is \textit{a priori} known by the receiver.
Given an estimate of $\mathbf{S}$ from the matrix factorization algorithm, we can
enhance the estimation accuracy of $\mathbf{G}$ (and hence $\mathbf{S}$) by
exploiting the fact that $\mathbf{G}$ is a sparse matrix (which is generally more sparse than $\mathbf{S}$).
This can be accomplished by using compressed sensing methods.

In the following, we propose a  semi-blind channel-and-signal estimation (SBCSE) approach.
The SBCSE algorithm {largely follows} the framework of BCSE, {except for} two extra steps.
In the first step, we use short pilots to estimate the phase and permutation
ambiguities. In the second step,
{we use  compressed sensing to improve the estimate of $\mathbf{G}$ after removing the phase and permutation ambiguities.}

\subsection{Estimation of Phase and Permutation Ambiguities}
We assume that the first $T_{\rm P}$ symbols of each user packet are short
pilots known by the receiver, i.e., $\mathbf{X}=[\mathbf{X}_{\rm P}, \mathbf{X}_{\rm D}]$,
where $\mathbf {X}_{\rm{P}}=[{\mathbf x}_{{\rm P},1},{\mathbf x}_{{\rm P},2},\cdots,{\mathbf x}_{{\rm P},KN}]^{T}\in \mathbb{C}^{KN\times T_{\rm P}}$ and data symbols
${\mathbf X}_{\rm D}=[{\mathbf x}_{{\rm D},1},\cdots,{\mathbf x}_{{\rm D},KN}]^{T}\in \mathbb{C}^{KN \times (T-T_{\rm P})}$.
Correspondingly, $\mathbf{Y}$ can be represented as $\mathbf{Y}=[\mathbf{Y}_{\rm P}, \mathbf{Y}_{\rm D}]$, where $\mathbf{Y}_{\rm P}\in \mathbb{C}^{M\times T_{\rm P}}$.
By ``short pilots", {we mean  $T_{\rm P}\ll KN$
so that the conventional training based channel estimation methods (including those based on compressed sensing)
cannot provide a good estimate of the channel solely based on $\mathbf{Y}_{\rm P}$ and $\mathbf{X}_{\rm P}$.
 Meanwhile, $T_{\rm P}\geq\lceil \log_{|{\cal A|}}KN\rceil+1$\footnote{\blue{For the $\log_{|A|}KN+1$ symbols, one symbol is used to eliminate the phase
 ambiguity and the remaining $\log_{|A|}KN$ symbols are used to eliminate the permutation
 ambiguity.}}, so that the phase and permutation ambiguities can be efficiently resolved.}

{The SBCSE algorithm is given as follows.} Let $\hat {\mathbf S}$ and $\hat {\mathbf{X}}$  be a pair of
output estimates from the inner iteration (at a certain round of outer iteration). Recall that
BCSE suffers from the phase and permutation ambiguities.
 Let $\bm{ \Pi} =[\bm{\pi}_1,\bm{\pi}_2,\cdots,{\bm \pi}_{KN}]^T\in  \mathbb {C}^{KN\times KN}$ be the  permutation ambiguity matrix,
where ${\bm \pi}_p$ is a $KN\times 1$  unit with only one non-zero entry. Likewise, let ${\bm \Sigma}={\rm diag}\{\sigma_1,\sigma_2,\cdots,\sigma_{KN}\}$
be the phase ambiguity matrix (with the diagonal elements being $\sigma_p=e^{j\varpi_p}$, $\varpi_p\in\Omega$, $p=1,\cdots,KN$) carried in $\hat {\mathbf{S}}$
 and $\hat {\mathbf{X}}$. Then, $\hat {\mathbf{S}}$ and $\hat {\mathbf{X}}$ can be written as
\begin{align}
\hat {\mathbf{S}}=\tilde {\mathbf{S}} {\bm {\Pi}}^{-1}{\bm {\Sigma}}^{-1} \;\text{and}\;  \hat {\mathbf{X}}=\bm{\Sigma}\bm{\Pi} \mathbf{X},
\end{align}
where $\tilde{\mathbf{S}}$ is the ambiguity-corrected estimate of $\mathbf{S}$.
Correspondingly, we write
\begin{align}\label{semi.pilot}
\hat {\mathbf{X}}_{\rm P}={\bm\Sigma}{\bm \Pi}\mathbf{X}_{\rm P},
\end{align}
where $\hat{\mathbf{X}}_{\rm P}$ consists of the first ${T}_{\rm P}$ columns of $\hat {\mathbf{X}}$.
Note that $\mathbf{X}_{\rm P}$ is known by the receiver. 

Recall from  Algorithm 1 that the distribution of $x_{p,t}$ at the end of a certain inner iteration is
given by $\tilde{p}(x_{p,t})$ in \eqref{post.x}.
As $x_{p,t}$ is discrete, for any given $p,t,k$, we denote by $P(\hat{x}_{p,t}=x_{k,t})$ the probability
of $\hat{x}_{p,t}=x_{k,t}$ specified by $\tilde{p}(x_{p,t})$ in \eqref{post.x}.
Let $\mathbf{u}_{k}$ be the $k$-th column of the $KN$-by-$KN$ identity matrix. Then, for any given $p$ and $t$,
the joint probability of $\sigma_{p}=e^{j\varpi_p}$ and $\bm{\pi}_p=\mathbf{u}_k$ is given by
\begin{align}
P(\sigma_{p}=e^{j\varpi_p},\bm{\pi}_p=\mathbf{u}_k)=\prod_{t=1}^{T_{\rm P}}P(\hat{x}_{p,t}=e^{j\varpi}x_{k,t}).
\end{align}
The corresponding marginals are given by
\begin{subequations}
\begin{align}
P(\sigma_p=e^{j\varpi})&=\sum_{k\in {\cal I}}\prod_{t=1}^{T_{\rm P}}P( \hat x_{p,t}=e^{j\varpi}x_{k,t}), \varpi\in \Omega  \label{semi.phase} \\
P(\bm{\pi}_p=\mathbf{u}_l)&=\sum_{w\in {\Omega}}\prod_{t=1}^{T_{\rm P}}P( \hat x_{p,t}=e^{j\varpi}x_{l,t}),l\in \{1,\cdots,KN\}\triangleq \cal I. \label{semi.permu}
\end{align}
\end{subequations}
Based on \eqref{semi.phase}, an estimate of $\bm{\Sigma}$ is given by
$\hat {\bm \Sigma}={\rm diag}\{{\hat \sigma}_1,{\hat \sigma}_2,\cdots,\hat{\sigma}_{KN}\}$,
where  $\hat {\sigma}_p=\arg\max_{\varpi \in \Omega}P(\sigma_p=e^{j\varpi})$.
Similarly, an estimate of $\hat{\bm \Pi}$ is given by
$\hat{\bm \Pi}=[\hat {\bm \pi}_1,\hat{\bm \pi}_2,\cdots,\hat{\bm\pi}_{KN}]$,
where $\hat {\bm\pi}_p=\arg\max_{l\in {\cal I}}P({\bm{\pi}_p=\mathbf{u}}_l)$.
\begin{algorithm}
\scriptsize
	\caption{\label{alg2}: SBCSE algorithm}
	\begin{algorithmic}[1]
	\State \textbf{Input:} received signal $\mathbf{Y}$, parameters  $\bm{\psi}\triangleq \{{\bm \vartheta},\rho,\lambda_{{S}},p_{01}^{{S}},p_{10}^{{S}},v_{{S}},\sigma^2\}$, prior distributions $p(x_{p,t})$ and $p(s_{q,p})$.		
		\State\textbf{Initialization:} ${\hat s}_{q,p}(1)={\hat w}_{q,t}(1)=0$, $ v_{q,p}^{s}(1)=v_{q,t}^{w}(1)=v_{\rm max}$, $\hat {x}_{p,t}(1)$  randomly chosen from ${\cal A}$, $v_{p,t}^x(1)=v_{\rm max}$, $\hat {\tau}_{m,t}(0)=\hat{\alpha}_{q,t}(0)=0$, $\forall q,t,p$
			
		\For{$i=1,2,3,\cdots, I_{\rm max}$ } \quad\quad\quad\quad{$ \textbf{\%}$} \textbf{outer iteration}
\For{$j=1,2,3,\cdots, J_{\rm max}$ }\quad\quad\quad\quad{$ \textbf{\%}$} \textbf{inner iteration}	
         \State Run lines 5-29 of Algorithm 1
        \EndFor
\State $\forall p$: $P(\sigma_p=e^{j\varpi})=\frac{1}{|\cal A|}\sum_{k\in {\cal I}}\prod_{t=1}^{T_{\rm P}}P(\hat x_{p,t}=e^{j\varpi}x_{k,t})$, $\varpi\in \Omega$
\State $\forall p$: $P(\bm{\pi}_p={\mathbf{u}}_{l})=\frac{1}{|\cal A|}\sum_{\varpi\in {\Omega}\prod_{t=1}^{T_{\rm P}}}P(\hat x_{p,t}=e^{j\varpi}x_{l,t})$, $l\in {\cal I}$
\State $\forall p$: $\hat {\sigma}_p=\arg\max_{\varpi \in \Omega}P(\sigma_p=e^{j\varpi})$;\;$\hat {\bm\pi}_p=\arg\max_{l\in {\cal I}}P(\bm{\pi}_p=\mathbf{u}_l)$
\State $\hat {\bm \Sigma}={\rm diag}\{\hat{\sigma}_1, \hat{\sigma}_2,\cdots,\hat{\sigma}_{KN}\}$;\; $\hat{\bm \Pi}=[\hat {\bm \pi}_1,\hat{\bm\pi}_2,\cdots,\hat{\bm\pi}_{KN}]$
\State \blue{$\tilde {\mathbf{S}}^{'}=\hat{\mathbf{S}}\hat {\bm \Sigma} \hat{\bm \Pi}$;\; $\bm{\varphi}=\bm{\varphi}^{1}$; \; $[{\hat{\mathbf{g}}}_1^{1},\cdots,{\hat{\mathbf{g}}}_{M^{'}}^{1}]=\mathbf{A}_{\textsf{T}}^H(\bm{\varphi}^{1})(\tilde{\mathbf{S}}^{'})^{H}$ }
\For {$t=1,2,3,\cdots, Q_{\rm max}$}
\State  \blue{$\forall k:$ $\mathbf{v}^t={\hat{\mathbf{g}}}_k^{t}+\frac{1}{\beta}(\mathbf{A}_{\textsf{T}}^H(\bm{\varphi}^t)((\tilde{\mathbf {S}}^{'})^H-\mathbf{A}_{\textsf{T}}(\bm{\varphi}^t){\hat{\mathbf{g}}}_k^{t}))$}
\State \blue{$\forall k:$ $\hat{{\mathbf{g}}}_k^{t+1}={\rm soft}(\mathbf{v}^t,{\alpha})$ }
\State \blue{Update $\bm{\varphi}^{t+1}$  via \eqref{parameter.tune}}
\EndFor
\State $\hat {{\mathbf G}}=[\hat{{\mathbf{g}}}_1^{t+1},\cdots,\hat{{\mathbf{g}}}_{M^{'}}^{t+1}]^{H}$;\;$\hat{\mathbf{S}}=\hat {\mathbf{G}}\mathbf{A}_{\textsf{T}}^H(\bm{\varphi}^{t+1})\hat{\bm \Pi}^{-1}\hat {\bm \Sigma}^{-1}$
\State {\bf Re-initialize} $\hat x_{p,t}(1)$, $v_{p,t}^x(1)$, $\hat s_{q,p}(1)$, and $v_{q,p}^s(1)$, $\forall q,p,t$
        \EndFor
\State $\tilde {\mathbf X}=\hat{{\bm \Pi}}^{-1}\hat{{\bm \Sigma}}^{-1}\hat {\mathbf X}$;\;$\tilde {\mathbf S}=\hat {\mathbf S}\hat{{\bm \Sigma}}\hat{\bm \Pi}$
\State\textbf{Output:} $\tilde {\mathbf X}$;\;$\tilde {\mathbf S}$
	\end{algorithmic}
\end{algorithm}

\subsection{Estimation of $\mathbf{G}$}

In this subsection, we further exploit the channel sparsity of ${\mathbf{G}}$ to enhance the channel estimate.
With  $\hat {\bm \Sigma}$ and $\hat{\bm \Pi}$, we  eliminate the ambiguities in $\hat {\mathbf{S}}$ as
\begin{align}
\tilde {\mathbf{S}}^{'}&=\hat{\mathbf{S}}\hat {\bm \Sigma} \hat{\bm \Pi}.
\end{align}
Recall  ${\mathbf {S}}=\mathbf{G}\mathbf{A}_{\textsf{T}}^H(\bm{\varphi})$ in \eqref{Y.propose.3}.  We model $\tilde {\mathbf{S}}^{'}$
as
\begin{align}\label{S_prime}
\tilde {\mathbf{S}}^{'}&=\mathbf {S} + \mathbf{W}^{'}\notag \\
&=\mathbf{G}\mathbf{A}_{\textsf{T}}^H(\bm{\varphi})+\mathbf{W}^{'},
\end{align}
where $\mathbf{W}^{'}$ is the additive noise contained in $\tilde{\mathbf{S}}^{'}$.
\blue{We aim to recover $\mathbf{G}$ from $\tilde{\mathbf{S}}^{'}$. To alleviate the possible angular mismatch for the
AoDs, we propose to  tune the angle grid at the user by considering the following optimization problem:}
\begin{align}\label{S_prime.tune}
\arg\min_{\bm{\varphi},\mathbf{G}}\|\tilde {\mathbf{S}}^{'}-\mathbf{G}\mathbf{A}_{\textsf{T}}^H(\bm{\varphi})\|_2^2+\alpha\big\|\mathbf{W}^{'}\big\|_1.
\end{align}
where $\alpha>0$ is a regularization factor.
To solve this problem, we  alternately update the estimates of ${\mathbf{G}}$ and ${\bm{\varphi}}$.
First, we aim to recover ${\mathbf{G}}$ from $\tilde{\mathbf {S}}^{'}$ for given ${\bm{\varphi}}$.
{Compressed sensing techniques can be used for this purpose. Since here the probability model of $\mathbf{W}^{'}$ is difficult to
acquire,} we propose to use the iterative soft-thresholding algorithm to deal with the sparsity\cite{wright2009sparse},
which is a robust estimator without requiring much knowledge of the statistical information of $\mathbf{W}^{'}$.
For the model in \eqref{S_prime}, the iterative soft thresholding algorithm is given by
{
\begin{align}\label{CS.2}
\mathbf{v}^t&={\hat{\mathbf{g}}}_k^{t}+\frac{1}{\beta}(\mathbf{A}_{\textsf{T}}^H(\bm{\varphi}^{t})((\tilde{\mathbf {S}}^{'})^H-\mathbf{A}_{\textsf{T}}(\bm{\varphi}^{t}){\hat{\mathbf{g}}}_k^{t})) \notag \\
\hat{{\mathbf{g}}}_k^{t+1}&={\rm soft}(\mathbf{v}^t,{\alpha}),
\end{align}}
where $t$ is the iteration number, $\beta$ is the maximum eigenvalue of {$\mathbf{A}_{\textsf{T}}^H(\bm{\varphi}^{t})\mathbf{A}_{\textsf{T}}(\bm{\varphi}^{t})$}, and ${\rm soft}(u,b)\equiv \frac{{\max \{|u|-b,0\}}}{\max\{|u|-b,0\}+b}u$. Note that in \eqref{CS.2}, ${\rm soft}(\cdot)$ is applied to vector $\mathbf{v}^t$ in a pointwise manner. \blue{Then, we aim
to improve the resolution of AoDs, i.e., to reduce the mismatch
between $\bm{\varphi}$ and true AoDs. We develop a gradient
descent method to solve \eqref{S_prime.tune} for given $\tilde {\mathbf{S}}^{'}$. Specifically, we compute
\begin{align}\label{parameter.tune}
\bm{\varphi}^{t+1}=\bm{\varphi}^{t}-\epsilon\frac{\bar{\bm{\varphi}}^t}{|\bar{\bm{\varphi}}^t|},
\end{align}
where $\bar{\bm{\varphi}}^t$ denotes the derivative of the objective in \eqref{S_prime.tune} with respect to $\bm{\varphi}$, and
$\epsilon$ is an appropriate  step size. }

The SBCSE algorithm is summarized in Algorithm 2.  Lines 4-6 compute the signal estimate $\hat {\mathbf{X}}$
and the channel estimate $\hat{\mathbf{S}}$ based on BCSE. In lines 7-11, we use short pilots to estimate the phase and permutation ambiguities, i.e., $\hat {\bm \Sigma}$ and $\hat {\bm \Pi}$. In lines 12-17, we  remove the phase and permutation ambiguities in $\hat{\mathbf{S}}$,  the compressed
sensing technique is used to improve the estimate of $\mathbf{G}$, and { a gradient descent method is used to improve the resolution of AoDs.} Line 20
 eliminates the phase and permutation ambiguities in $\hat{\mathbf{X}}$ and $\hat{\mathbf{S}}$. 

 \section{Further Discussions}
\subsection{Parameter Learning}
Recall from \eqref{post.prob} that the model parameters $\bm {\psi}$ are assumed to be known  by
the receiver. In practice, most of these parameters are unknown and need to be estimated.
We now describe an expectation maximization (EM) based
approach for parameter learning \cite{parker2014bilinear,liu2018super}. Specifically, at the $i$-th outer iteration,
we have the following updating rules for the parameters in $\bm{\psi}$:\footnote{Signal sparsity $\rho$ is not updated in \eqref{em.learning}, since $\rho$ is determined
by the transmission protocol.  In addition, $p_{10}^{{S}}$
is not included in \eqref{em.learning} since it can be updated by using the equality $\lambda_{S}=\frac{p_{01}^{S}}{p_{01}^{S}+p_{10}^{S}}$ once
$\lambda_{S}$ and $p_{01}^{S}$ are determined.}
\begin{subequations}\label{em.learning}
\begin{align}
\sigma^2(i+1)&=\arg\max_{\sigma^2}\mathbb{E}\left[\ln p(\mathbf{Y},\mathbf{S},\mathbf{X};\bm{\vartheta}(i),\rho(i),\lambda_{{S}}(i),p_{01}^{S}(i),v_{S}(i),\sigma^2)\right] \\
\vartheta_q(i+1)&=\arg\max_{\vartheta_q}\mathbb{E}\left[\ln p(\mathbf{Y},\mathbf{S},\mathbf{X};\vartheta_{1}(i+1),\cdots,\vartheta_{q-1}(i+1),\right. \notag \\
&\quad\quad\quad\left.\vartheta_{q},\vartheta_{q+1}(i),\cdots,\vartheta_{M^{'}}(i),\rho(i),\lambda^{S}(i),p_{01}^{S}(i),v_{S}(i),\sigma^2(i+1))\right] \\
v_{{S}}(i+1)&=\arg\max_{v_{S}}\mathbb{E}\left[\ln p(\mathbf{Y},\mathbf{S},\mathbf{X};\bm{\vartheta}(i+1),  \rho(i),\lambda_{{S}}(i),p_{01}^{S}(i),v_{S}, \sigma^2(i+1))\right]
\end{align}
\begin{align}
p_{01}^{S}(i+1)&=\arg\max_{p_{01}^{S}}\mathbb{E}\left[\ln p(\mathbf{Y},\mathbf{S},\mathbf{X};\bm{\vartheta}(i+1),  \rho(i),\lambda_{{S}}(i),p_{01}^{S},v_{S}(i+1), \sigma^2(i+1))\right]\\
\!\!\!\lambda_{S}(i+1)&=\arg\max_{\lambda_{S}}\mathbb{E}\!\!\left[\ln p(\mathbf{Y},\mathbf{S},\mathbf{X};\!\bm{\vartheta}(i+1),  \rho(i),\lambda_{{S}},p_{01}^{S}(i+1),v_{S}(i+1), \sigma^2(i+1))\!\right]\!\!\!\!\!
\end{align}
\end{subequations}
where the expectations in \eqref{em.learning} are taken over the distribution of $p(\mathbf{S},\mathbf{X}|\mathbf{Y};\bm{\psi}(i))$. However, the exact form of $p(\mathbf{S},\mathbf{X}|\mathbf{Y};\bm{\psi}(i))$ is difficult to
obtain. {Here,} we approximate the joint posterior distribution $p(\mathbf{S},\mathbf{X}|\mathbf{Y};\bm{\psi}(i))$
by the product of its marginals, i.e.,
\begin{align}
p(\mathbf{S},\mathbf{X}|\mathbf{Y};\bm{\psi}(i))&=\left(\prod_{q}\prod_{p}p(s_{q,p}|\mathbf{Y};\bm{\psi}(i))\right)\left(\prod_{t}\prod_{p}p(x_{p,t}|\mathbf{Y};\bm{\psi}(i))\right),
\end{align}
where {$p(x_{p,t}|\mathbf{Y};\bm{\psi}(i))$ and $p(s_{q,p}|\mathbf{Y};\bm{\psi}(i))$ at the $i$-th outer iteration are approximated   by \eqref{post.x} and \eqref{post.s}, respectively.}

\subsection{Complexity Analysis}
We now compare the computational complexity of various sparse matrix factorization methods for the system models
in \eqref{input.output.sparse} and \eqref{Y.propose.1}-\eqref{Y.propose.3}. We first consider the P-BiGAMP algorithm for
\eqref{input.output.sparse}. Recall from \eqref{vec.Y} that the size of ${\rm vec}({\mathbf{G}})\in \mathbb{C}^{KN^{'}M^{'}\times 1}$ is $KN^{'}M^{'}\times 1$,
that of ${\rm vec}({\mathbf{X}})\in \mathbb{C}^{KNT\times 1}$ is $KNT\times 1$,
and that of $\mathbf{z}_{l}$ is $MT \times 1$.  From \cite{parker2016parametric}, the complexity of P-BiGAMP is ${\cal O}(I_{\rm max}J_{\rm max}MT^2NK^2M^{'}N^{'})$,
which is prohibitive highly for massive MIMO systems. {Note that $I_{\rm max}$ and $J_{\rm max}$ are the maximum
numbers of outer iterations and inner iterations, respectively.} Second, we consider the matrix factorization problems
in \eqref{Y.H} and \eqref{Y.propose.1}, which can be solved by the BiGAMP algorithm. From \cite{parker2014bilinear},
the computational complexity is ${\cal{O}}(I_{\rm max}J_{\rm max}MTKN)$. Third, we consider the matrix factorization
problem in \eqref{Y.propose.3}. For BCSE in Algorithm 1, the complexity of lines 5-8 is ${\cal{O}}(MM^{'}T)$;
the complexity of lines 9-17 is ${\cal O}(M^{'}TKN)$; the complexity of lines 18-25
 is ${\cal{O}}(M^{¡®})$. Thus, the overall complexity is ${\cal{O}}(I_{\rm max}J_{\rm max}(M^{'}TKN+MM^{'}T))$.
{For SBCSE in Algorithm 2, the extra steps in lines 12-15 require complexity of ${\cal O}(K^2NN^{'}Q_{\rm max})$},
where $Q_{\rm max}$ is the maximum number of iterations for iterative soft thresholding. Thus,
the overall complexity of SBCSE is ${\cal O}(I_{\rm max}(J_{\rm max}M^{'}TKN+J_{\rm max}MM^{'}T+K^2NN^{'}Q_{\rm max}))$.
Finally, we consider the model in \eqref{Y.propose.2}. Note that
 the BCSE algorithm can be straightforwardly applied to \eqref{Y.propose.2}, except
that the prior distributions of the elements of $\tilde {\mathbf X}$ are replaced by Gaussian distributions.
 The involved complexity is ${\cal{O}}(I_{\rm max}J_{\rm max}(M^{'}TKN^{'}+MM^{'}T))$.
Also note that the SBCSE algorithm cannot be applied to \eqref{Y.propose.2}, since it is difficult to estimate
the phase and permutation ambiguities in factorizing $\mathbf{G}$ and $\tilde{\mathbf{X}}$
by using the pilots. The above discussions are summarized in Table II.

\begin{small}
\begin{table}[htbp]
	\centering
	\caption{Computational complexity}
	\label{tab3}
	\begin{tabular}{cc}
		\toprule
        Method & Complexity\\
		\hline
		P-BiGAMP for \eqref{Y.H.b} & ${\cal O}(I_{\rm max}J_{\rm max}MT^2NK^2M^{'}N^{'})$ \\
		BiGAMP for \eqref{Y.H} and \eqref{Y.propose.1}&${\cal{O}}(I_{\rm max}J_{\rm max}MTKN)$ \\
		BCSE for \eqref{Y.propose.2} & ${\cal{O}}(I_{\rm max}J_{\rm max}(M^{'}TKN^{'}+MM^{'}T))$\\
BCSE for \eqref{Y.propose.3} & ${\cal{O}}(I_{\rm max}J_{\rm max}(M^{'}TKN+MM^{'}T))$\\
SBCSE for \eqref{Y.propose.3} & ${\cal O}(I_{\rm max}(J_{\rm max}M^{'}TKN+J_{\rm max}MM^{'}T+K^2NN^{'}Q_{\rm max}))$\\
		\hline
	\end{tabular}
\end{table}
\end{small}
\section{Numerical Results}
In this section, we present simulation results to evaluate the performance of the BCSE and SBCSE algorithms.
In the simulations,  quadrature phase shift keying (QPSK) modulation with Gray-mapping is employed.
 The signal-to-noise ratio (SNR) is defined as $\frac{\rho KN}{\sigma^2}$.
For both  BCSE and SBCSE algorithms, the maximum number of inner iterations $L_{\rm max}$ is set to 200,
and the maximum number of outer iterations $M_{\rm max} $ is set to 20. The number of random initializations is
set to 5.

We are now ready to compare the performance of different approaches, as listed below.
\begin{itemize}
\item BiGAMP for \eqref{Y.H}: $\mathbf{H}$ and $\mathbf{X}$ in \eqref{Y.H} are recovered using the BiGAMP algorithm by exploiting the signal sparsity \cite{ding2018sparsity}.
  \item BiGAMP for \eqref{Y.propose.1}: $\mathbf{S}$ and $\mathbf{X}$ in \eqref{Y.propose.1} are recovered using the BiGAMP algorithm.
  \bluenew{Both the sparsity of $\mathbf{S}$ and the sparsity of $\mathbf{X}$ are exploited in the algorithm.
  Note that the sparsity level $\rho$ of $\mathbf{X}$  is known by the algorithm, whereas the sparsity level $\lambda_{S}$ of
  $\mathbf{S}$ is learned using the EM method described in Section V-A. }
  \item {Training-based: In the training-based scheme, we use the BiGAMP algorithm for
sparse matrix factorization, with the signal means and variances are initialized (and re-initialized)
by following the strategy in \cite{wen2016bayes}. That is, the first $T_{\rm P}$
columns of $\mathbf{X}$ are initialized as the pilots $\mathbf{X}_{\rm P}$,
and the corresponding variances are set to zero.} 
  \item BCSE for \eqref{Y.propose.2}: The proposed blind detection scheme (Algorithm 1) is applied to the factorization of $\mathbf{G}\tilde{\mathbf{X}}$
in  \eqref{Y.propose.2}, except that the prior distributions of the elements of $\tilde{\mathbf{X}}$
are replaced by Gaussian distributions.
  \item BCSE for \eqref{Y.propose.3}: The proposed blind detection scheme   (Algorithm 1) is applied to the factorization of $\mathbf{S}\mathbf{X}$ in \eqref{Y.propose.3}.
  \item SBCSE for \eqref{Y.propose.3}: The proposed semi-blind detection scheme   (Algorithm 2)  is applied to the factorization of $\mathbf{S}\mathbf{X}$ in \eqref{Y.propose.3}.
  \item \bluenew{Genie bound with $\mathbf{S}$ known: The proposed blind detection scheme (Algorithm 1) is applied to the factorization
of $\mathbf{S}$ and $\mathbf{X}$ in \eqref{Y.propose.3}, where $\mathbf{S}$ is known to the receiver.}
  \item \bluenew{Genie bound with $\mathbf{X}$ known: The proposed blind detection scheme (Algorithm 1) is applied to the factorization
of $\mathbf{S}$ and $\mathbf{X}$ in \eqref{Y.propose.3}, where $\mathbf{X}$ is known to the receiver.}
\end{itemize}

We use the bit-error rate (BER) of the signal and the normalized mean square error (NMSE) of the channel as the evaluation metrics. All the simulation results are obtained by taking  {average} over 100 random realizations.

\subsection{Blind Channel-and-Signal Estimation}
\begin{figure}[htbp]
  \centering
  \subfigure[BER of $\mathbf{X}$]{\includegraphics[width=2.5 in]{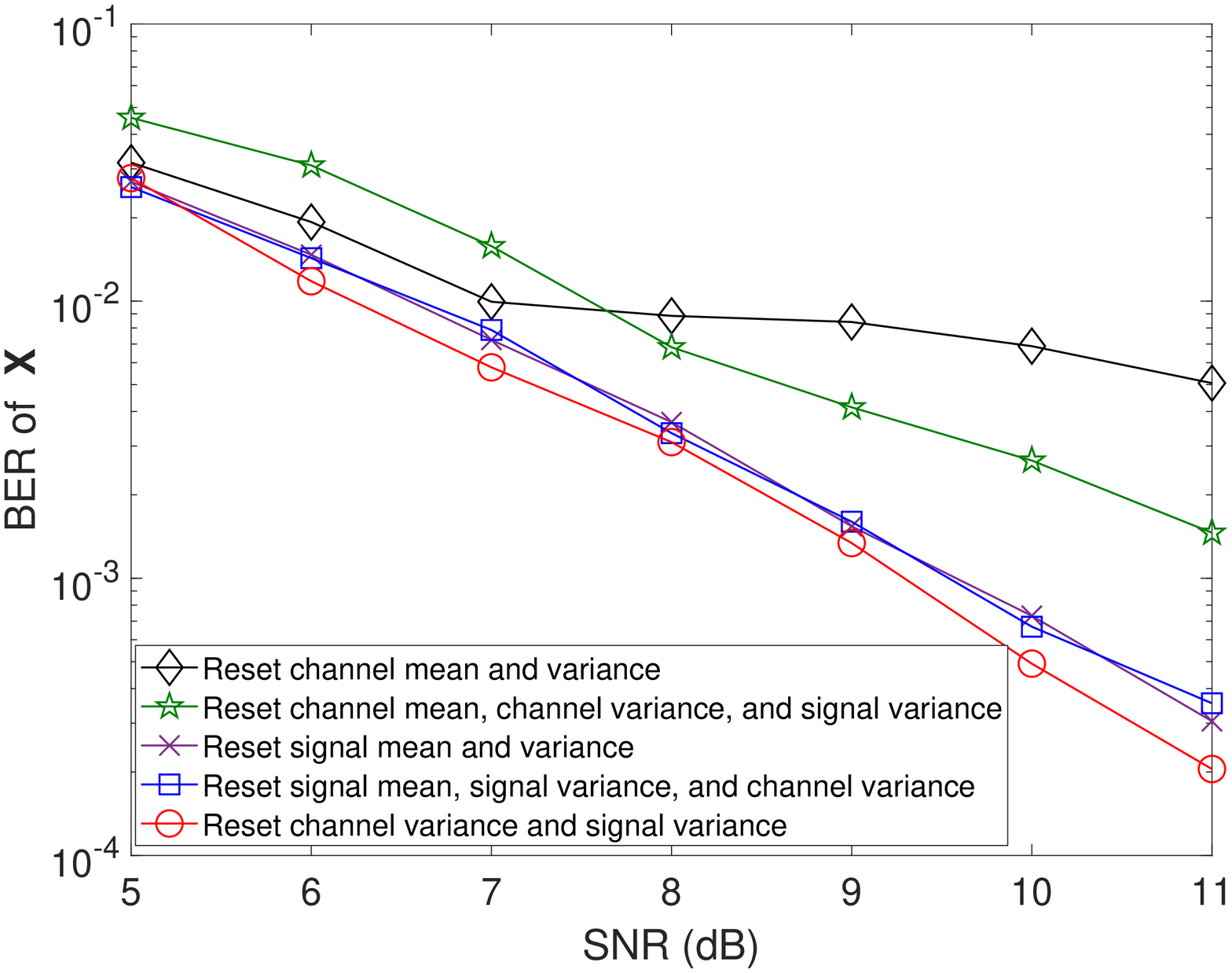}}
  \subfigure[NMSE of $\mathbf{H}$]{\includegraphics[width=2.5 in]{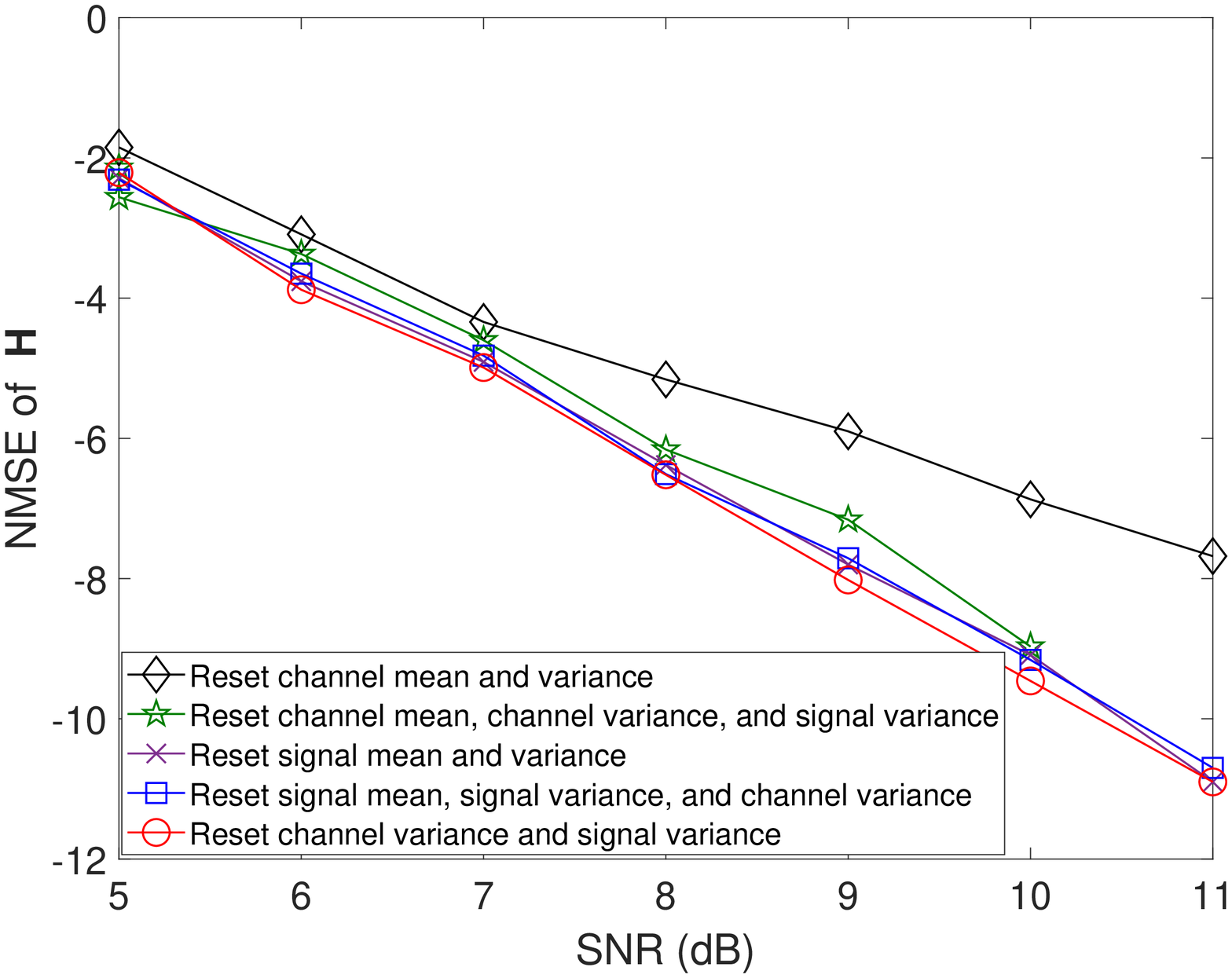}}
  \caption{{The performance comparison of different re-initialization methods, where $M=M^{'}=128$, $K=20$, $N=N^{'}=1$, $T=100$, $\rho=0.1$, and $\lambda=0.1$.}}\label{fig.initial}
\end{figure}

 In the simulations, the true AoAs are generated by
\begin{align}\label{AoAs}
\sin(\theta_q)&=\sin(\vartheta_q^{\rm DFT})+\varsigma_q, \text{with}\; \varsigma_q \sim {\rm U}\left[-\frac{1}{2M^{'}},\frac{1}{2M^{'}}\right]
\end{align}
where $\{\vartheta_q^{\rm DFT}\}$  is the DFT sampling grids, and ${\rm U}[c,d]$ denotes the uniform distribution
over $[c,d]$. We assume that the receiver exactly knows the true AoAs, so that the grid $\{\vartheta_{q}\}_{q=1}^{M^{'}}$
used at the receiver perfectly covers the true AoAs $\{\theta_{q}\}_{q=1}^{M^{'}}$.
The entries of $\mathbf{G}$ are randomly and independently drawn from the distribution \blue{$p(g_{k,q,p})=(1-\lambda)\delta(g_{k,q,p})+\lambda {\cal C}{\cal N}\left(g_{k,q,p};0,v_{\rm pri}\right)$} with
 $\lambda =0.1$ and $v_{\rm pri}=1$. Other parameters are set as $M=M^{'}=128$, $N=N^{'}=1$, $K=20$, $T=100$, $\rho=0.1$, and $v_{\rm max}=10$.
Fig.~\ref{fig.initial} shows the BER and NMSE performance of the five different  re-initialization methods (discussed in Section III-D) versus SNR.
We see that the method by resetting the channel mean and the channel variance (used in \cite{liu2018super}) has a
BER error floor at around $10^{-2}$. Further resetting the signal variance does not work well either.
The other three resetting strategies have better performance.
Among them, the best performance occurs when only the signal variance and the channel variance are reset.
Hence, we always use this re-initialization method in the remaining simulation results.
\begin{figure}[htbp]
  \centering
  \subfigure[$\rho=0.4$]{\includegraphics[width=2.5 in]{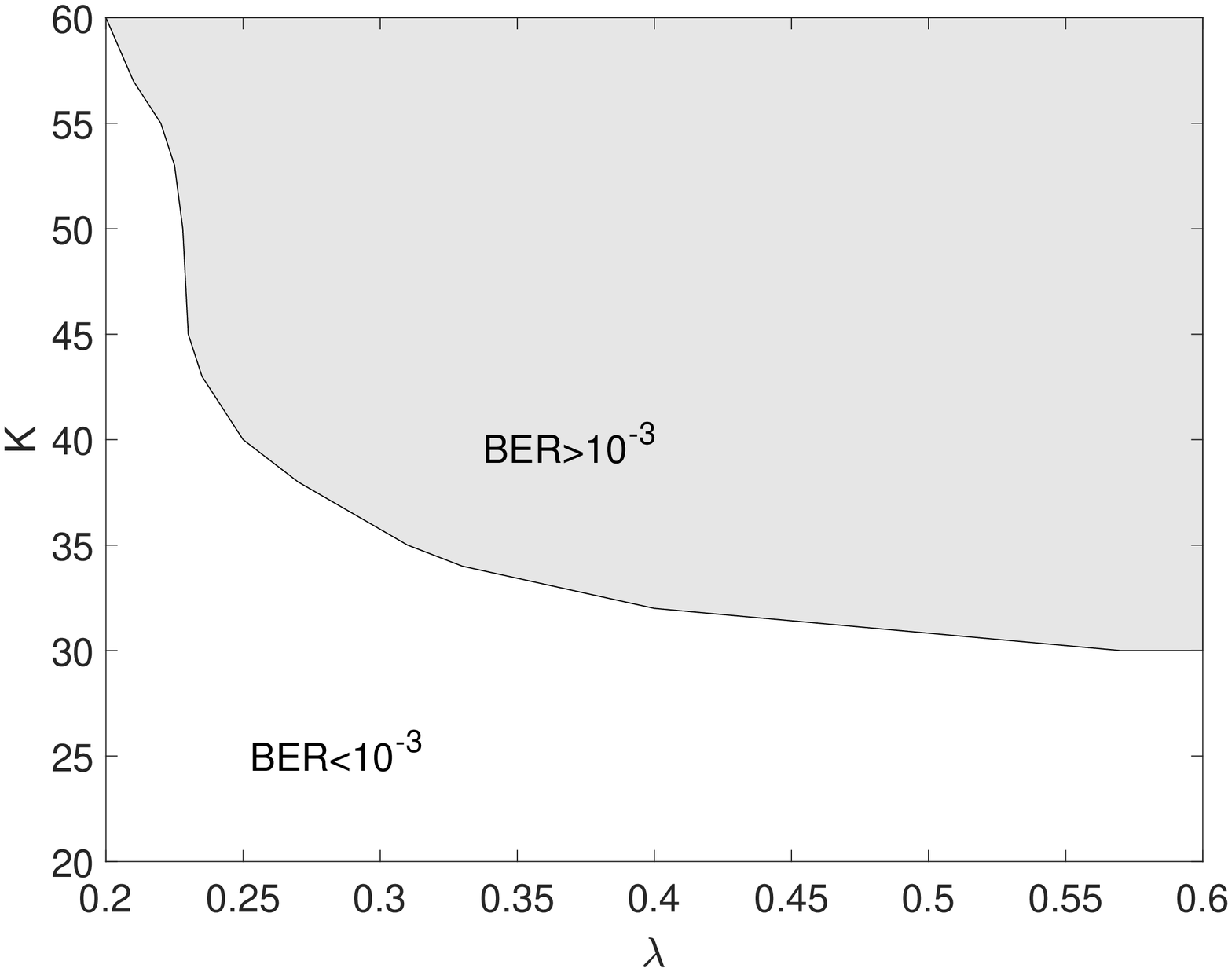}}
  \subfigure[$\lambda=0.4$]{\includegraphics[width=2.45 in]{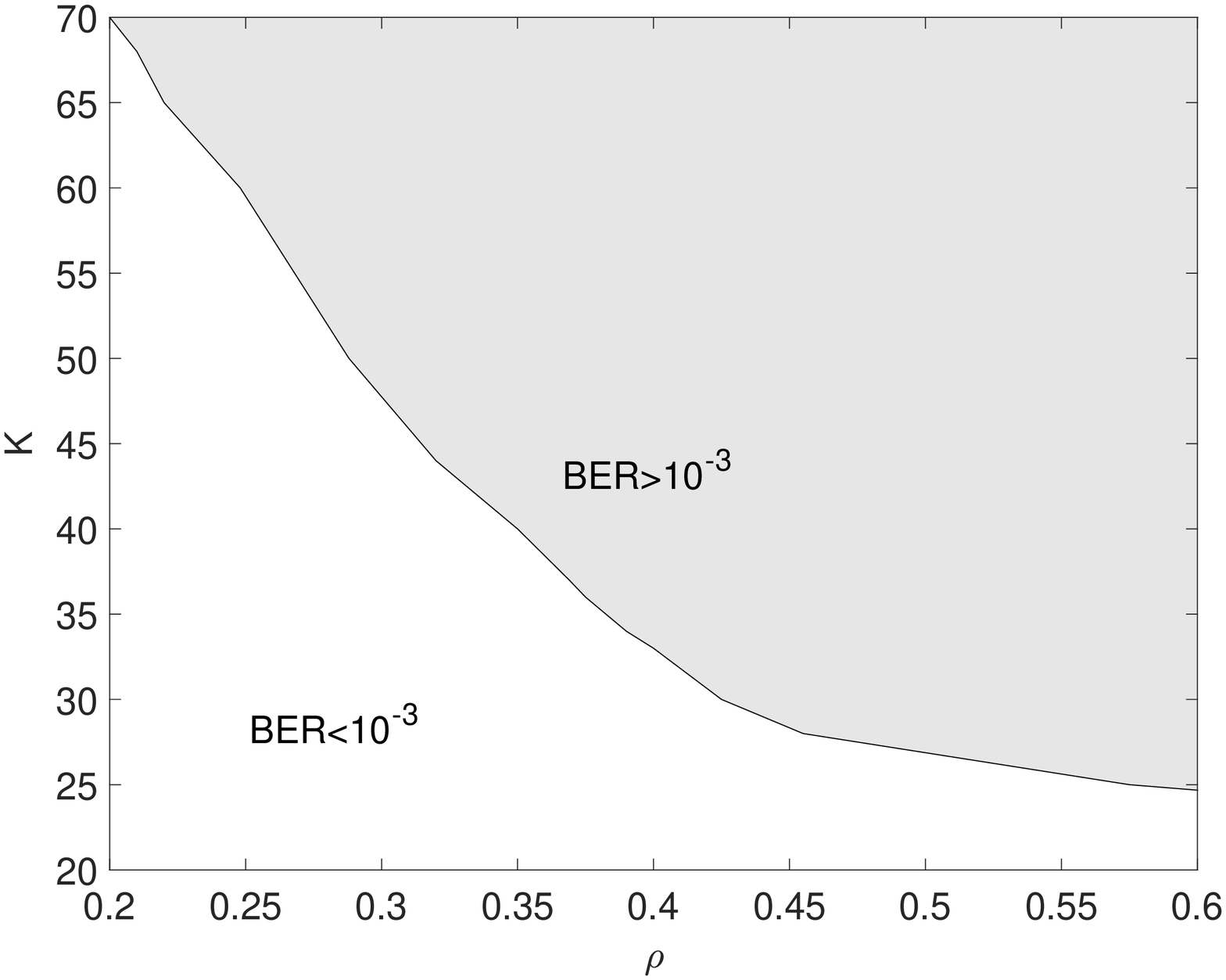}}
  \caption{The phase transition of the BCSE scheme for the model in \eqref{Y.propose.3}, where $M=M^{'}=128$, $N=N^{'}=1$, and $T=256$.}\label{fig.threhod.ratio}
\end{figure}

Fig.~\ref{fig.threhod.ratio} shows the requirements on $K$, $\rho$, and $\lambda$
for successful recovery with other parameters fixed at $M=M^{'}=128$, $N=N^{'}=1$,  $T=256$, and SNR $=20$ dB.
We say that the recovery is successful and the corresponding value of $K$, $\rho$, and $\lambda$ are feasible if the BER of $\mathbf{X}<10^{-3}$.
Fig.~\ref{fig.threhod.ratio}(a)  shows the feasiable region of $(K,\lambda)$ with the signal sparsity $\rho=0.4$.
 We see that the boundary is  a monotonic function of the sparsity level $\lambda$, i.e., the sparser the channel is,
the greater the number of users can be supported. It is also interesting to
 see that the system is able to perform successful recovery with $K=30$, even for
a relatively large $\lambda$. In this case, the user signals are still separable due
to the signal sparsity. Fig.~\ref{fig.threhod.ratio}(b) shows the available region of $(K,\rho)$ with the channel sparsity $\lambda=0.4$.
We  observe that the boundary is also a  monotonic function of the sparsity level $\rho$, as expected.

\begin{figure}[]
  \centering
\subfigure[BER of $\mathbf{X}$]{\includegraphics[width=2.35 in]{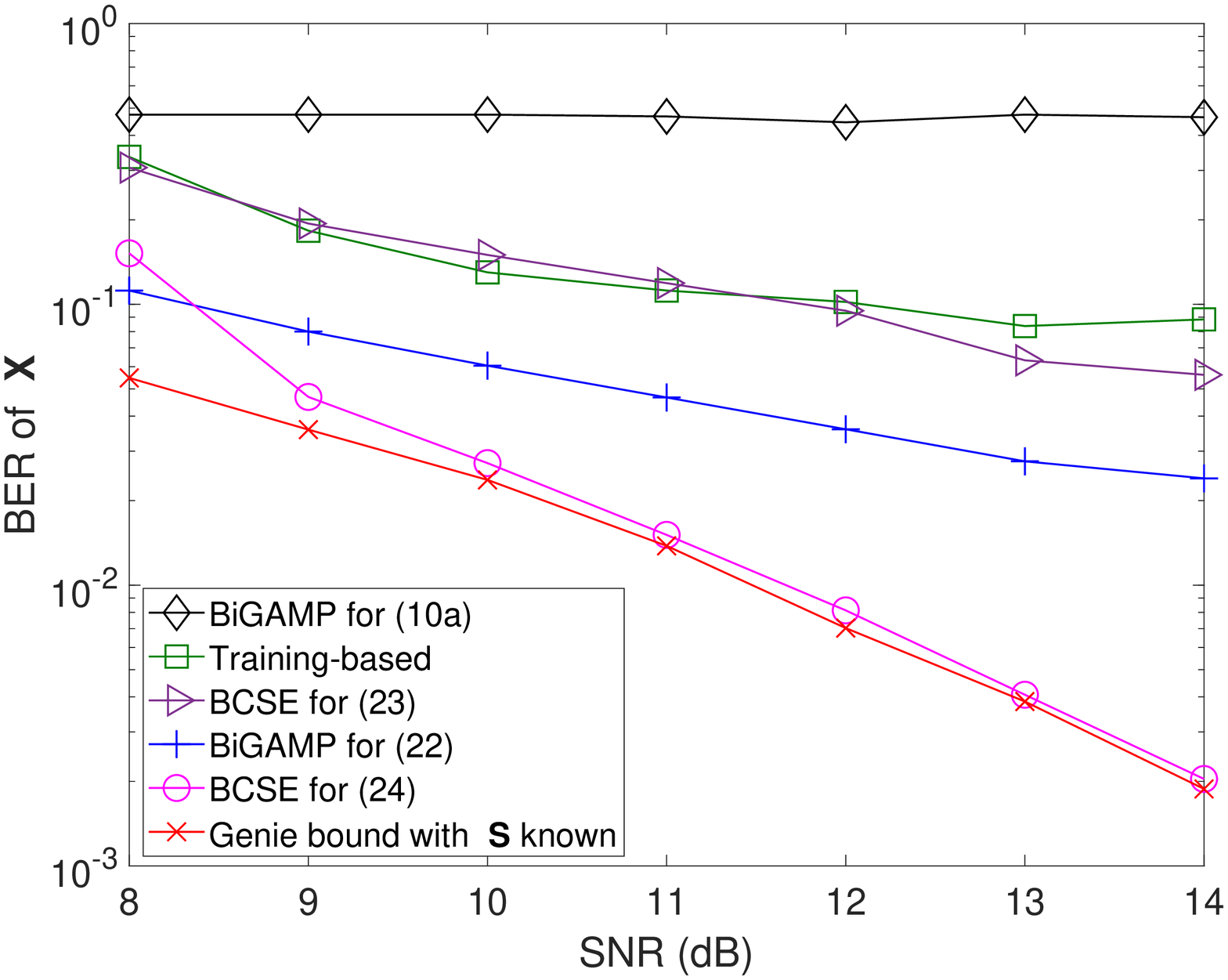}}
\subfigure[NMSE of $\mathbf{H}$]{\includegraphics[width=2.35 in]{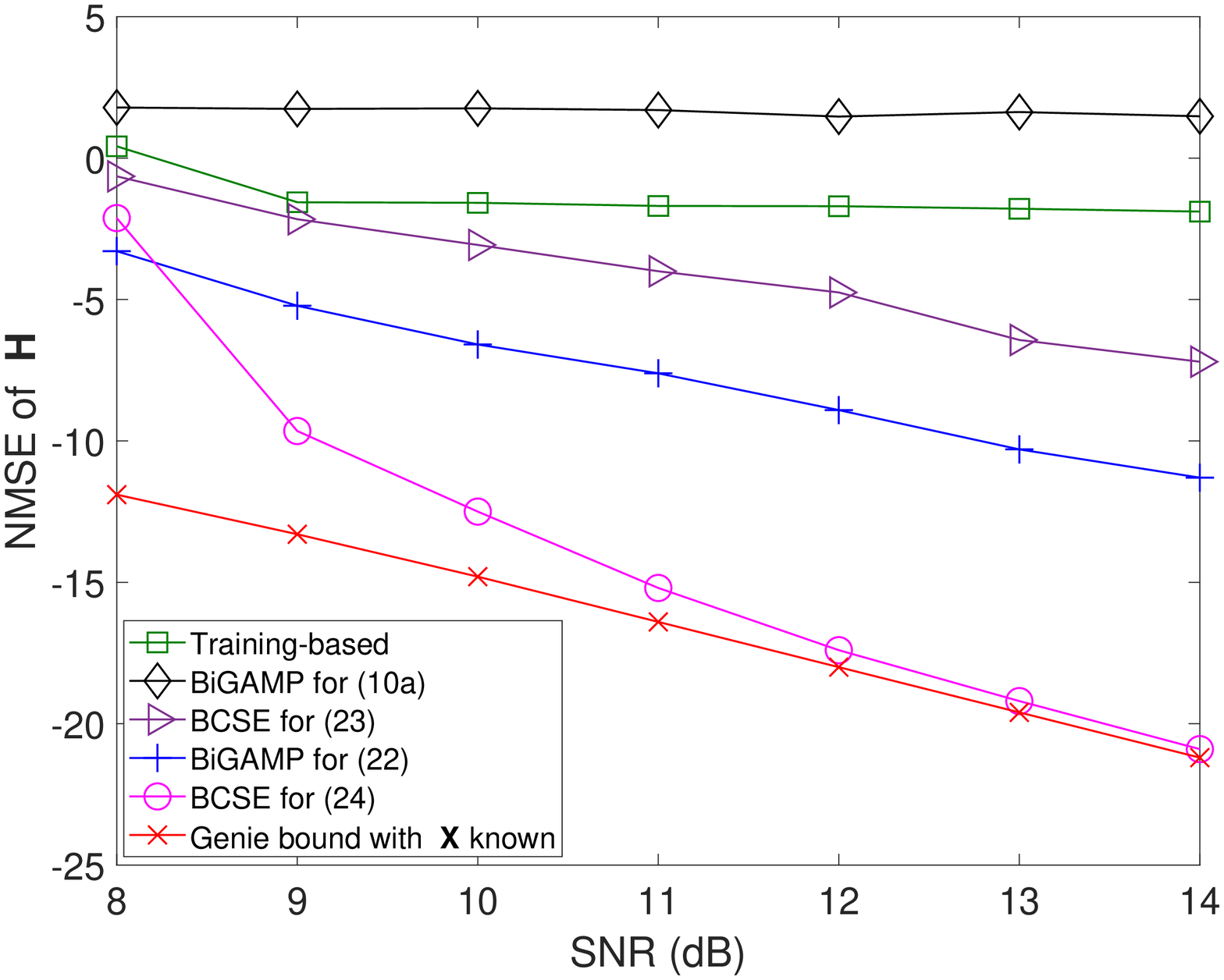}}
  \caption{$M=M^{'}=128$, $N=N^{'}=1$, $K=30$, $T=256$, $\rho=0.4$, and $\lambda=\lambda_{S}=0.1$.}\label{fig.blind}
\end{figure}

Fig.~\ref{fig.blind} shows the blind detection performance versus  SNR, where $M=M^{'}=128$, $N=N^{'}=1$, $K=30$, $T=256$, $\rho=0.4$, and $\lambda=\lambda_{S}=0.1$.
$T_{\rm P}=1+\lceil \log_430\rceil$ QPSK symbols are used as pilots for each packet.
For blind detection schemes, these pilots are used to remove phase and permutation ambiguities.
From Fig.~\ref{fig.blind},  we see that the BiG-AMP method  only exploiting the signal sparsity in \eqref{Y.H} does not work well,
neither does the training-based method. The BCSE for \eqref{Y.propose.2} has obvious performance loss than the BCSE
for \eqref{Y.propose.3} since a loss of constellation constraints.  The BiGAMP for \eqref{Y.propose.1} suffers from
an error floor due to the unavoidable energy leakage problem of using the DFT basis for grid sampling.
Clearly, our proposed BCSE algorithm significantly outperforms the counterpart schemes. \bluenew{Also, the proposed
BCSE algorithm approaches the genie bound in the high SNR regime.} 

\begin{figure}[]
  \centering
 \includegraphics[width=2.5 in]{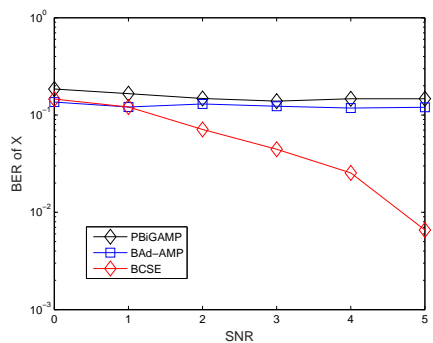}
  \caption{BER Performance of different schemes in a small setup, where $M\!=\!M^{'}\!=\!32$, $K=1$, $N\!=\!N^{'}\!=\!4$, $T\!=\!50$, $\rho\!=\!0.2$, and $\lambda\!=\!0.2$.}\label{fig.small.scale}
\end{figure}

\blue{The proposed BCSE algorithm can also compare with the other bilinear recovery algorithm, such as  BAd-VAMP in \cite{BAdAMP19} and PBIGAMP in \cite{parker2016parametric}.
However, these algorithms will have high complexity in the settings in Fig.~\ref{fig.blind}, so we try to compare in a small setup.  We reset  simulation  settings  as  $M=M^{'}=32$, $N=N^{'}=4$, $K=1$, $T=50$, $\rho=0.2$, and $\lambda=0.2$.
 From Fig.~\ref{fig.small.scale}, we see that BAd-VAMP and PBiGAMP do not work well. The
proposed BCSE algorithm considerably  outperforms  BAd-VAMP and PBiGAMP.}
\subsection{Performance of Semi-Blind Channel-and-Signal Estimation}
\begin{figure}[]
  \centering
 \includegraphics[width=2.7 in]{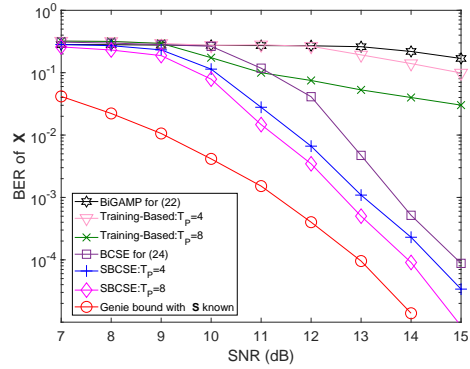}
  \caption{BER Performance of different schemes with $M\!=\!M^{'}\!=\!128$, $K=3$, $N\!=\!N^{'}\!=\!8$, $T\!=\!100$, $\rho\!=\!0.3$, and $\lambda\!=\!0.1$.}\label{fig.semi.blind}
\end{figure}

We still assume that the true AoAs generated by \eqref{AoAs} and the AoA
grid perfectly covers the true AoAs.
 A uniform sampling grid is adopted to generate the AoDs, i.e., $\mathbf{A}_{\textsf{T,k}}(\bm \varphi_k)$
is a DFT matrix for each user $k$. The   parameter settings are $M=M^{'}=128$, $N=N^{'}=8$, $K=3$, $T=100$, $\rho=0.3$, and $\lambda=0.1$.
Fig.~\ref{fig.semi.blind}  shows the BER performance
versus SNR for various schemes with the number of pilots $T_{\rm P}=$ 4 and 8.
From Fig.~\ref{fig.semi.blind},
we see that the SBCSE method for $T_{\rm P}=4$ significantly outperforms the training-based scheme and BiGAMP for \eqref{Y.propose.1}.
The SBCSE scheme outperforms
the BCSE scheme by about 1 dB at BER $=10^{-4}$ for $T_{\rm P}=8$.  We
also see that, as in contrast to Fig.~\ref{fig.blind}, there is a performance gap of about 1 dB between the SBCSE scheme and the genie bound with $\mathbf{S}$ known
in the relatively high SNR regime. This is caused by the suboptimal estimation for $\mathbf{G}$ using the iterative soft-thresholding algorithm.

\begin{figure}[]
  \centering
 \includegraphics[width=2.7 in]{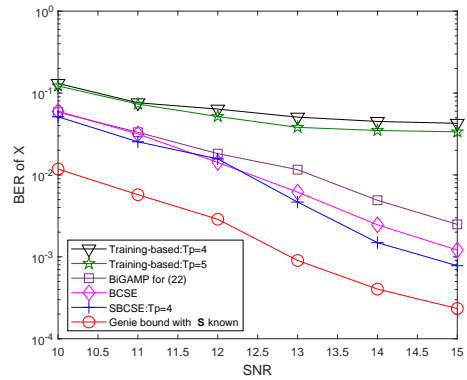}
  \caption{\blue{BER Performance of different schemes under LAA, where $M\!=\!M^{'}\!=\!128$, $K=3$, $N\!=\!N^{'}\!=\!8$, $T\!=\!100$, $\rho\!=\!0.3$, and $\lambda\!=\!0.1$.}}\label{fig.LAA.performance}
\end{figure}

\blue{We extend  our algorithms to the LAA antenna array, with the corresponding steering vectors given by
\begin{small}
\begin{align}
\mathbf{a}_\textsf{R}(\theta_k)&=\left[\textrm{sinc}\!\left(-\frac{M^{'}-1}{2}-\frac{D}{\varrho}\sin \theta_k \! \right),\textrm{sinc}\!\left(-\frac{M^{'}-3}{2}-\frac{D}{\varrho}\sin \theta_k\!\right),\cdots,\textrm{sinc}\left(\frac{M^{'}-1}{2}-\frac{D}{\varrho}\sin \theta_k\right)\right]^{T}\notag \\
\mathbf{a}_\textsf{T}(\phi_{k})&=\left[\textrm{sinc}\left(-\frac{N^{'}-1}{2}-\frac{D}{\varrho}\sin \phi_{k}\right),\textrm{sinc}\left(-\frac{N^{'}-3}{2}-\frac{D}{\varrho}\sin \phi_{k}\right),\cdots,\textrm{sinc}\left(\frac{N^{'}-1}{2}-\frac{D}{\varrho}\sin \phi_{k}\right)\right]^{T},
\end{align}
\end{small}
where $\textrm{sinc}(\cdot)$ denotes the nominalized ``sinc'' function,
and $D$ denotes the lens length along the azimuth plane.
We have added the simulation results in the LAA antenna geometry in Fig.~\ref{fig.LAA.performance}.
We see a similar performance trend in Fig.~\ref{fig.LAA.performance} as the case of ULA in Fig.~\ref{fig.semi.blind}.}

\begin{figure}[]
  \centering
  \includegraphics[width=2.7 in]{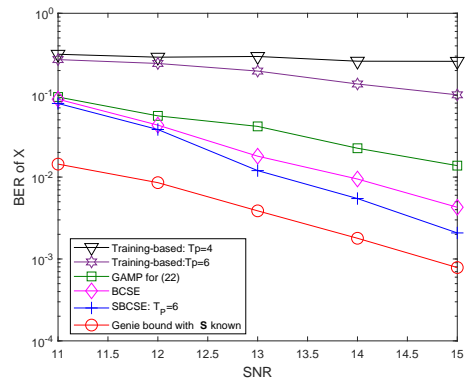}
  \caption{\blue{BER Performance of different schemes in large-scale fading, where $M\!=\!M^{'}\!=\!128$, $K=3$, $N\!=\!N^{'}\!=\!8$, $T\!=\!100$, $\rho\!=\!0.3$, and $\lambda\!=\!0.1$.}}\label{fig.semi.blind.largescale} \end{figure}

\blue{We further study the impact of large-scale fading on the system
performance. The channel powers $v_{\rm pri}$ of the $k$-th user are randomly drawn
from a uniform distribution over $[v_{\rm pri,\rm min},1]$. Figure.~\ref{fig.semi.blind.largescale} shows the performance of the various schemes
in the presence of large-scale fading. In simulations, we set $-10\log10(v_{\rm pri,\rm min})=20$ dB, and
the EM algorithm  in Section V is employed for the tuning of of $v_{\rm pri}$. The other
settings are the same as those in Fig.~\ref{fig.semi.blind}. From Fig.~\ref{fig.semi.blind.largescale},
we see that the trends of the curves are very similar to those in Fig.~\ref{fig.semi.blind}.}

\begin{figure}[]
  \centering
 \subfigure[BCSE]{\includegraphics[width=2.5 in]{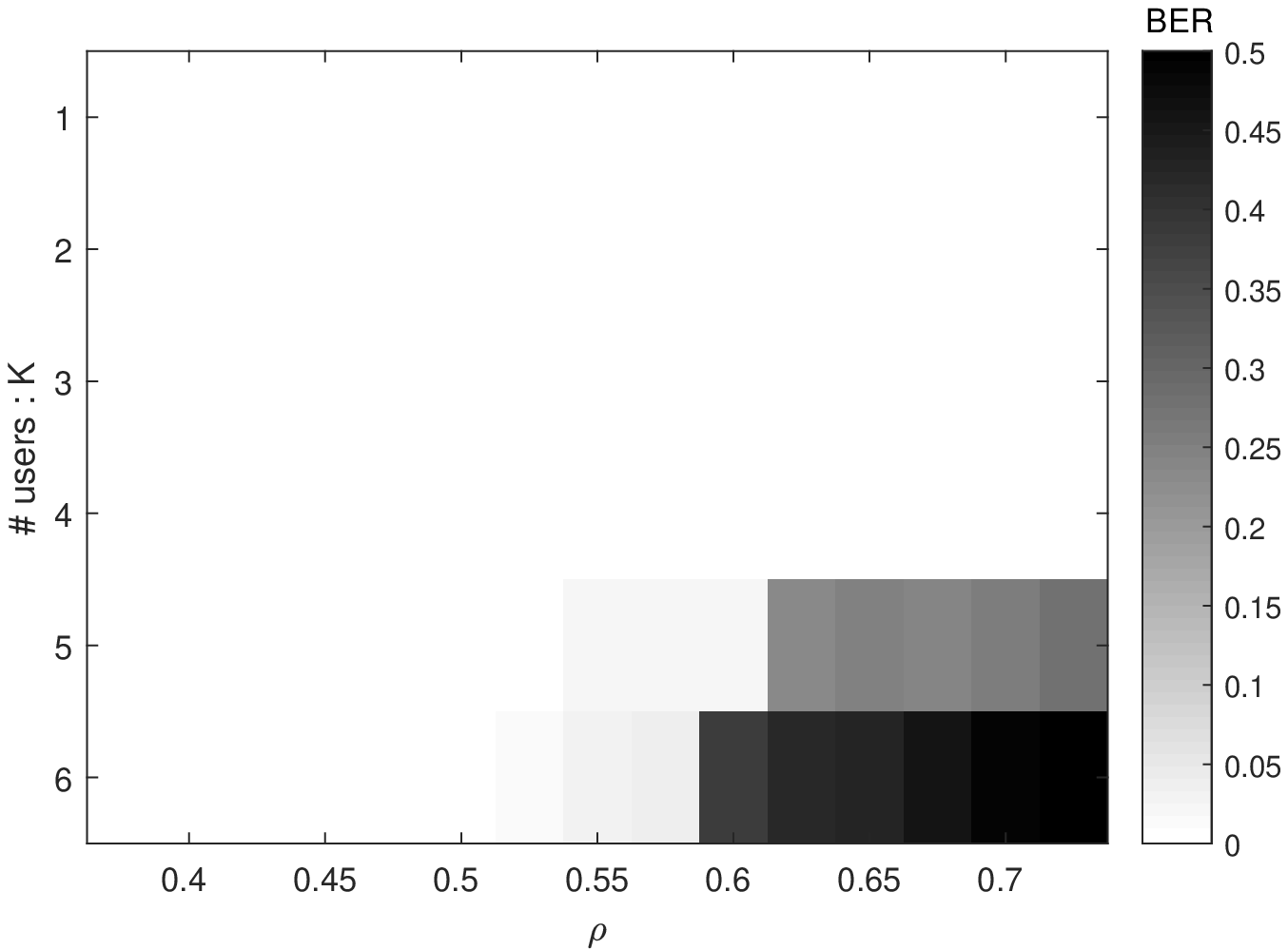}}
\subfigure[SBCSE]{\includegraphics[width=2.5 in]{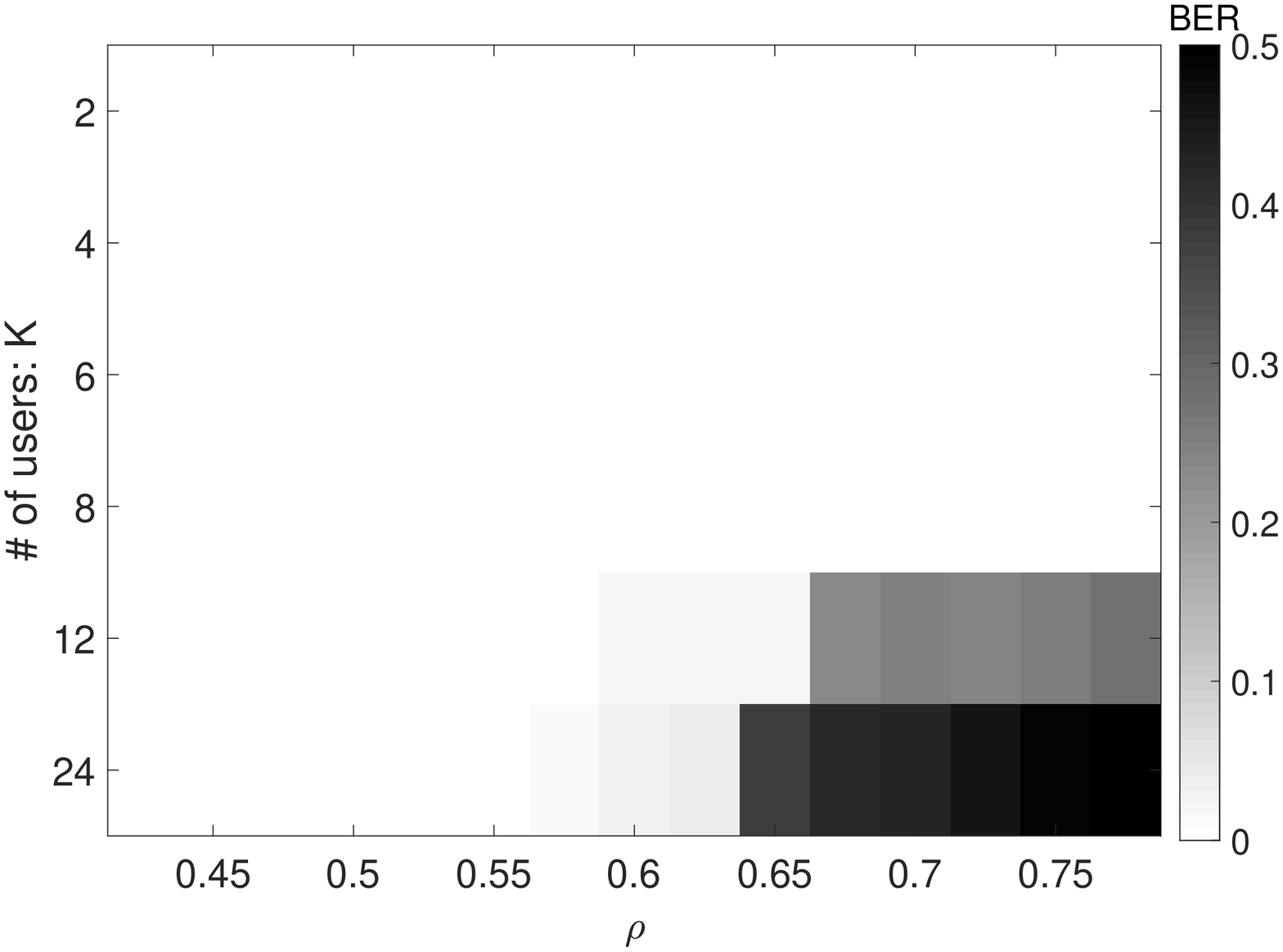}}\\
  \caption{Phase transition diagrams for BCSE and SBCSE without parameter tuning, where $M=M^{'}=128$, $KN=24$, $T=100$, $T_{\rm P}=8$, and SNR $=20$ dB.}\label{fig.transition}
\end{figure}

 Fig.~\ref{fig.transition} shows the transition diagrams for the
BCSE and SBCSE schemes, where $M=M^{'}=128$, $T=100$, $T_{\rm P}=8$, $KN=24$, $N=N^{'}$, $\lambda=0.1$, and SNR $=20$ dB.
Clearly, for fixed $KN$, $\lambda_{S}$ increases monotonically with $N$, or in other words, decreases with $K$.\footnote{
In simulation, the value of $\lambda_{S}$ is obtained by using the parameter learning technique described in Section V-A.}
We see SBCSE works well in a much broader region of the channel and the signal sparsity than BCSE does.

\begin{figure}[]
  \centering
 \includegraphics[width=2.7 in]{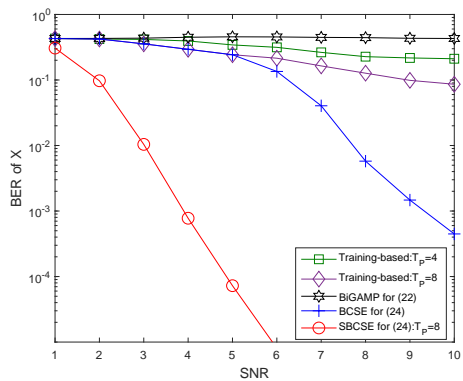}
  \caption{Performance of different schemes with parameter tuning, where $M=M^{'}=128$, $T=100$, $T_{\rm P}=8$, $K=3$, $N=N^{'}=8$, and $\rho=0.35$.}\label{fig.para.tune}
\end{figure}
\subsection{BCSE and SBCSE  with {Parameter} Learning}
 We now assume that the AoAs are unknown to the BS.
The channel is generated by the model in \eqref{channel.phy}.
Specially, we generate the center angle of each scattering
cluster uniformly from $[-\frac \pi 2, \frac \pi 2]$,
 and the AoA of each subpath $\theta(i,j)$ concentrates in a $20^{\circ}$ angular spread.
$L_{1,c}=\cdots=L_{K,c}=3$ and $L_{1,p}=\cdots=L_{K,p}=50$.
For the AoDs, we assume that $\{\phi_k(i,j)\}$ fall on  a uniform sampling grid in the virtual angular domain.
Fig.~\ref{fig.para.tune}  shows the BER performance against  SNR
with $M=M^{'}=128$, $T=100$, $T_{\rm P}=8$, $K=3$, $N=N^{'}=8$, and $\rho=0.35$.
The Markov chain model in \eqref{markove.S} is used for characterising  the
clustering effect of the channel.
The parameters in $\bm{\psi}$ are tuned using the EM method in Section V-A.
Similar trends as in Fig.~\ref{fig.semi.blind} has been observed in Fig.~\ref{fig.para.tune}.
Particularly, the SNR gap between SBCSE ($T_{\rm P}=8$) and BCSE is enlarged
to over $5$ dB at BER $=10^{-3}$.

\begin{figure}[]
\setlength{\belowcaptionskip}{-1cm}
  \centering
 \includegraphics[width=2.5 in]{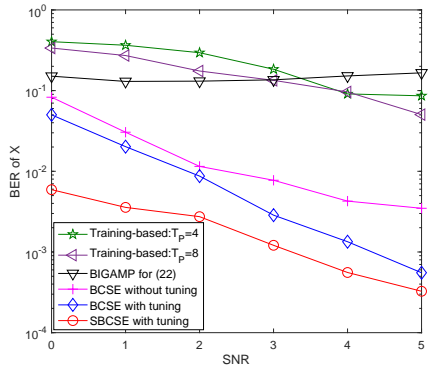}
  \caption{BER Performance of different schemes in SCM channel model,  where $M\!=\!M^{'}\!=\!128$, $K=1$, $N\!=\!N^{'}\!=\!8$, $T\!=\!100$, $\rho\!=\!0.4$.}\label{fig.semi.blind.SCM}
\end{figure}

\blue{We further  add the experiment in  the  spatial channel model (SCM) developed
in 3GPP/3GPP2 for low frequency band (less than 6 GHz) \cite{WinnerScmImplementationIEEETranbst}.
The parameters of SCM used in the simulations are listed in Table.~\ref{tab1}.
Fig.~\ref{fig.semi.blind.SCM} shows the BER performance of different schemes against  SNR
with $M=M^{'}=128$, $T=100$, $T_{\rm P}=8$, $K=1$, $N=N^{'}=8$, and $\rho=0.4$.
\begin{table}[htbp]
	\centering
	\caption{Parameter Settings for the Channel Model}
	\label{tab1}
	\begin{tabular}{cccc}
		\toprule
		\multicolumn{4}{c}{Parameter Settings for the SCM} \\
		\hline
		Parameter name & Value & Parameter name & Value \\
		\hline
        Scenario &`urban$\_$macro'& CenterFrequency & 2GHz\\
		NumBsElements   & 128    & NumMsElements & 8\\
		 NumPaths & 3 &  NumSubPathsPerPath        & 20  \\
		\hline
	\end{tabular}
\end{table}
From Fig.~\ref{fig.semi.blind.SCM}, we see that the BiGAMP method does not work well, neither  does the training-based method.
The BCSE with AoAs and AoDs tuning outperforms the BCSE without angles tuning. Further, we see that the
SBCSE scheme with AoAs and AoDs tuning considerably outperforms the BCSE scheme for $T_{\rm P}$= 8.}

\section{Conclusions}
In this paper, we have studied joint antenna activity detection, channel estimation,
and multiuser detection for massive MIMO system
with GSM. We first designed the BCSE algorithm by exploiting the double-sparsity of the system model.
We further developed the SBCSE algorithm,
 where a {short pilot sequence} is first used to estimate the phase and permutation ambiguities
 and then compressed sensing  is adopted to enhance the estimation performance.
 Extensive numerical results have been provided to demonstrate the superior performance of
 the proposed BCSE and SBCSE algorithms over
 the state-of-the-art blind detection and training-based
algorithms.



\end{document}